\documentclass[5p, authoryear, twocolumn]{elsarticle}   	
\usepackage{geometry}                		
\usepackage{graphicx}				
\usepackage{amssymb}

\usepackage{subfigure}

\usepackage{wrapfig}
\usepackage{lscape}
\usepackage{rotating}
\usepackage{epstopdf}
\usepackage{afterpage}

 \usepackage{mathtools}
 \usepackage{bm}
 
\usepackage{csvsimple}
\usepackage{longtable}
\usepackage{booktabs}
\usepackage{array}

\usepackage{pdflscape}
 
 \usepackage{subfig}
 
 \usepackage{chapterbib}

\usepackage{footmisc}
 
 \usepackage[switch]{lineno}

\newcommand*\patchAmsMathEnvironmentForLineno[1]{%
  \expandafter\let\csname old#1\expandafter\endcsname\csname #1\endcsname
  \expandafter\let\csname oldend#1\expandafter\endcsname\csname end#1\endcsname
  \renewenvironment{#1}%
     {\linenomath\csname old#1\endcsname}%
     {\csname oldend#1\endcsname\endlinenomath}}%
\newcommand*\patchBothAmsMathEnvironmentsForLineno[1]{%
  \patchAmsMathEnvironmentForLineno{#1}%
  \patchAmsMathEnvironmentForLineno{#1*}}%
\AtBeginDocument{%
\patchBothAmsMathEnvironmentsForLineno{equation}%
\patchBothAmsMathEnvironmentsForLineno{align}%
\patchBothAmsMathEnvironmentsForLineno{flalign}%
\patchBothAmsMathEnvironmentsForLineno{alignat}%
\patchBothAmsMathEnvironmentsForLineno{gather}%
\patchBothAmsMathEnvironmentsForLineno{multline}%
}

\usepackage{natbib}

\begin{document}\sloppy

\clearpage
\clearpage

\begin{frontmatter}

\author[cal,harv]{Simon J. Lock \corref{cor}}
\author[dav]{Sarah T. Stewart}
\author[seti]{Matija \'{C}uk}
\cortext[cor]{Corresponding author: slock@caltech.edu}

\address[cal]{Division of Geological and Planetary Sciences, California Institute of Technology, 1200 E. California Blvd., Pasadena, CA 91125, U.S.A.}
\address[harv]{Department of Earth and Planetary Sciences, Harvard University, 20 Oxford Street, Cambridge, MA 02138, U.S.A.}
\address[dav]{Department of Earth and Planetary Sciences, University of California Davis, One Shields Avenue, Davis, CA 95616, U.S.A.}
\address[seti]{SETI Institute, 189 Bernardo Avenue, Mountain View, CA 94043, USA }

\title{The energy budget and figure of Earth during recovery from the Moon-forming giant impact}


\begin{abstract} 

Quantifying the energy budget of Earth in the first few million years following the Moon-forming giant impact is vital to understanding Earth's initial thermal state and the dynamics of lunar tidal evolution. After the impact, the body was substantially vaporized and rotating rapidly, very different from the planet we know today. The subsequent evolution of Earth's energy budget, as the body cooled and angular momentum was transferred during lunar tidal recession, has not been accurately calculated with all relevant energy components included. Here, we use giant impact simulations and planetary structure models to calculate the energy budget at stages in Earth's evolution. We show that the figure and internal structure of Earth changed substantially during its post-impact evolution and that changes in kinetic, potential, and internal energy were all significant. These changes have important implications for the dynamics of tidal recession and the thermal structure of early Earth. 

\vspace{20px}
\raggedright
\vspace{8px}
\newline
\noindent\copyright  2019. This manuscript version is made available under the CC-BY-NC-ND 4.0 license http://creativecommons.org/licenses/by-nc-nd/4.0/

\end{abstract}

\end{frontmatter}

\section{Introduction}
\label{sec:intro}

\begin{figure}
\centering
\includegraphics[]{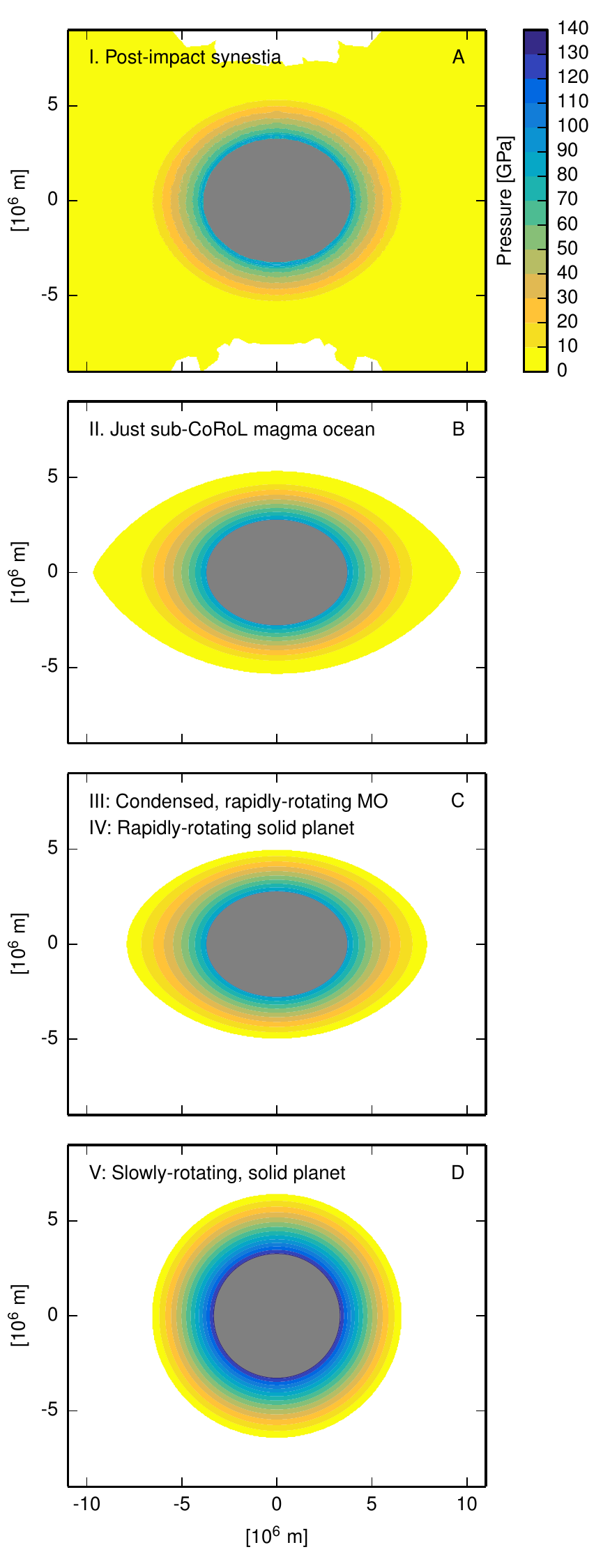}
\caption{Caption opposite.}
\end{figure}
\setcounter{figure}{0}
\begin{figure}
\caption{The shape and physical structure of the post-impact body changed dramatically after the Moon-forming giant impact. Shown are mantle pressure contours for a post-impact synestia (I), and for the same body once it has cooled to just below the corotation limit (II), cooled to a magma-ocean planet with a volatile-dominated atmosphere (III), the mantle has frozen (IV), and after the Moon has tidally receded (V) to the point at which the system undergoes the Cassini-state transition \citep[which occurred when the lunar semi-major axis was about 30 Earth radii,][]{Ward1975}, at which point Earth had an AM of 0.417~$L_{\rm EM}$. In this case, the post-impact body was formed by a collision between two $0.52 M_{\rm Earth}$ bodies at 9.7~km~s$^{-1}$ with an impact parameter of 0.55. Adapted from \citet{Lock2019pressure}.
}
\label{fig:pcontours}
\end{figure}

The last major event in Earth's accretion is thought to have been the Moon-forming giant impact \citep{Hartmann1975,Cameron1976}. Despite the giant-impact hypothesis being the favored lunar origin model for over 40 years, the parameters of the impact are still debated \citep[see recent reviews by][]{Asphaug2014,Barr2016}. \citet{Cameron1976} proposed that the impact prescribed the present-day angular momentum (AM) of the Earth-Moon system ($L_{\rm EM} = 3.5 \times 10^{34}$~kg~m$^{2}$~s$^{-1}$) and numerical simulations have shown that a modest-energy, grazing impact could satisfy this constraint and potentially produce a lunar-mass moon \citep{Canup2004}. This scenario, which we refer to as the canonical scenario, has become the de facto working model for lunar origin. However, recent work has cast doubt on the ability of the canonical model to reproduce key observations of the Earth-Moon system \citep{Melosh2014,Kruijer2017,Lock2018moon}. In recent years, several mechanisms have been found that could have reduced the AM of the system after the impact by three-body interactions between the Sun, Earth, and Moon \citep{Cuk2012,Wisdom2015,Cuk2016,Tian2017}. This discovery has dramatically increased the range of possible Moon-forming impacts and a variety of high-energy, high-AM collisions have been proposed \citep{Cuk2012,Canup2012,Lock2018moon}.

The Moon-forming impact set the stage for Earth's subsequent evolution. After all proposed impacts, the body would have been substantially vaporized and rotating rapidly \citep{Lock2017}. We use the term post-impact body to describe all bound mass, excluding any large satellites. Furthermore, in high-AM models the body immediately after the impact is not a planet, but a different type of planetary object, named a synestia \citep{Lock2017}. Synestias are bodies that exceed the corotation limit (CoRoL), a point defined by the AM at which the angular velocity at the equator of a corotating body (i.e., one that is in rigid-body rotation) equals that of a circular Keplerian orbit. The CoRoL is a function of thermal state, AM, total mass, and compositional layering. Subsequently, the body cooled and condensed, and its rotation rate was slowed by tidal recession of the Moon. 

Recovery after the Moon-forming impact is a key stage in Earth's evolution, but many facets of this period have not been studied. During recovery, the thermal structure and shape of the body changed substantially. Here, we use both shape and figure to describe the geometry of the body. \citet{Lock2019pressure} showed that internal pressures in the post-impact body could have been much lower than in present-day Earth (10s of GPa lower at the core-mantle boundary). The lower pressures are due to the lower density of the substantially-vaporized post-impact body, the rapid rotation of the body leading to a strong centrifugal force that counteracted gravity, and the bodies distorted shape. The pressures would have subsequently increased during cooling and lunar tidal recession as the body condensed and its rotation was slowed. Lower pressures in the aftermath of giant impacts change how the mantle freezes and provide a new paradigm for interpreting geochemical tracers of accretion.

Here, we consider another aspect of the recovery of Earth: the energy budget. The energy budget evolved rapidly as the body cooled and AM was transferred away from Earth. Constraining this evolution is vital for understanding processes that dominated early Earth. For example, the energy of the post-impact body dictates the time taken for the silicate vapor to condense and the body to transition to a magma-ocean planet. The mass of silicate vapor that persisted during satellite formation determined the balance between different processes in lunar accretion and controlled the Moon's composition \citep{Lock2018moon}. Furthermore, while the body was partially vaporized it had an increased collisional cross section and the rate at which impact debris was reaccreted was higher \citep[see discussion in][]{Lock2017}. 

Energy dissipation during lunar tidal recession heated the planet. Tidal recession reduces the Moon's orbital energy and the AM of Earth, liberating energy which was heterogeneously dissipated as heat within Earth and the Moon, perturbing their thermal profiles. The total amount of energy dissipated was determined by the initial AM of the system. The higher the initial AM of Earth, the greater the amount of energy that must have been lost to slow its rotation to its present-day period. Depending on the timing of recession, the energy dissipated could have controlled the timescale for magma ocean freezing and Earth's initial thermal structure.

The rate of recession, and hence tidal heating, was controlled by the tidal properties of Earth, including its figure. For scenarios where the initial AM of the Earth-Moon system is higher than the present-day, a variety of mechanisms have been proposed to transfer AM away from the system, including: the evection resonance \citep{Cuk2012}; an evection-based limit cycle \citep{Wisdom2015}; and an instability during the Laplace-plane transition for an initially high-obliquity Earth \citep{Cuk2016}. The tidal properties of a planet are highly sensitive to its thermal structure \citep[e.g.,][]{Henning2014}, and there is a feedback between the rate of tidal recession and the thermal state which could have controlled the orbital evolution of the Moon \citep{Zahnle2015}. Quantifying the tidal properties of Earth is necessary to calculate the rate of tidal recession, the efficiency of AM transfer, and thus the rate of change in Earth's energy budget.

Despite its importance, the energy budget of Earth after the impact has not been fully quantified. Previous work has typically considered Earth's evolution only after the majority of the silicate vapor had condensed, and neglected the changes in Earth's shape during cooling and tidal recession \citep[e.g.,][]{Peale1978,Elkins-Tanton2008,Lebrun2013,Zahnle2015}. Here, we calculate the evolution of Earth's energy budget starting immediately after the impact and include the effects of changes in shape. We also consider the influence of changes in figure on lunar tidal evolution and the rate of energy dissipation.

It is not yet feasible to construct a dynamical model of how Earth transitioned from a hot, rapidly rotating post-impact body to today's solid, slowly rotating planet. Here, we take the approach of comparing the energy budget at different stages during recovery: (I) immediately after the impact; (II) once the body has cooled to just below the CoRoL; (III) once the body has fully condensed to a magma ocean; (IV) after the mantle has frozen; and (V) during lunar tidal recession. Figure~\ref{fig:pcontours} shows pressure contours at each of these stages. At the first stage, a few dynamical times ($\sim48$~hrs) after the impact, the body is substantially vaporized, extended and rotating rapidly (Figure~\ref{fig:pcontours}A). The mantle transitions smoothly from vapor to supercritical fluid to liquid at high pressure, and there is no liquid surface overlain by a silicate atmosphere \citep{Stewart2018LPSC}. The body radiates at about 2300~K \citep{Lock2018moon}, driving rapid condensation of the silicate vapor. If the body was initially above the CoRoL, it would cool to below the limit and become a corotating planet (Figure~\ref{fig:pcontours}B, stage II). At this stage there is still a substantial mass of silicate vapor, but the body probably has a liquid surface \citep{Stewart2018LPSC}. At the third stage, most of the vapor has condensed and the body has a liquid upper mantle overlain by a volatile-dominated atmosphere (Figure~\ref{fig:pcontours}C). The body is substantially oblate. After some high-AM Moon-forming impacts, the planet would have an equatorial radius twice that of its polar radius. The planet continues to cool and solidifies over 10s~kyr \citep{Lebrun2013} to 10s~Myr \citep{Elkins-Tanton2008,Zahnle2015} (Figure~\ref{fig:pcontours}C, stage IV). The rate of tidal dissipation in a partially molten Earth is highly uncertain \citep[e.g.,][]{Henning2014,Zahnle2015} and so, for illustration, we consider the limiting case where the Moon remained close to Earth during magma-ocean solidification, and the AM did not change. As the Moon tidally recedes, the AM of Earth decreases, and its shape approaches spherical (Figure~\ref{fig:pcontours}D). We calculate the energy budget of Earth with different AM, i.e., at different points in its rotational evolution (stage V).

We calculate the energy budget at each stage using smoothed particle hydrodynamic simulations of giant impacts and a potential field method (Section~\ref{sec:methods}). We determine how each of the energy terms described in Section~\ref{sec:Eterms} change during condensation (Section~\ref{sec:results:cooling}) and tidal recession (Section~\ref{sec:results:tidal}). We then discuss the implications of our results for the cooling of Earth and lunar tidal recession (Section~\ref{sec:disc}) and conclude (Section~\ref{sec:conc}). Supplementary materials contain detailed methods and additional figures. 

\section{Overview of energy components}
\label{sec:Eterms}

We consider three principal energy components: kinetic, gravitational potential and internal. We describe each of these components and identify contributions to the energy budget during recovery after a giant impact that arise from significant changes in shape and which are not fully accounted for when assuming a constant moment of inertia or minimally-deformed figure, as has been done in previous work \citep[e.g.,][]{Peale1978,Zahnle2015}. Here we consider the energy of the equilibrium structure and rotation of bodies and do not include the contribution from small scale dynamics (e.g., convection) or non-equilibrium effects as such terms are typically much smaller.

Kinetic energy can be separated into two components: linear and rotational. Linear kinetic energy is significant for the impact, but we analyze the post-impact body in the center-of-mass reference frame in which the linear component is zero. The rotational kinetic energy is given by
\begin{align}
\begin{aligned}
E_{\rm K} = \frac{1}{2}\int_\mathcal{V}{r_{xy}^2 \omega^2 \rho \mathrm{d}V'} \, ,
\end{aligned}
\label{eqn:KE_base}
\end{align}
where $r_{xy}$ is distance from the rotation axis, $\omega$ is angular velocity, $\rho$ is density, $V$ is volume, and $\mathcal{V}$ is the total volume of the body. The rotation rate in post-impact bodies varies with distance from the rotation axis and so it is necessary to use the full expression above to calculate the rotational kinetic energy, but for corotating bodies, Equation~\ref{eqn:KE_base} reduces to
\begin{align}
\begin{aligned}
E_{\rm K, co} = \frac{1}{2} I \omega^2 \, ,
\end{aligned}
\label{eqn:KE_corot_omg}
\end{align}
where $I$ is the moment of inertia of the body. In terms of the body's AM, $L=I \omega$,
\begin{align}
\begin{aligned}
E_{\rm K, co} =  \frac{L^2}{2I}  \, .
\end{aligned}
\label{eqn:KE_corot_L}
\end{align}
The kinetic energy evolves due to changes in the moment of inertia (i.e., the distribution of mass) or AM. Under the constant moment of inertia approximation, kinetic energy only changes due to AM evolution.

The gravitational binding energy of a body is given by
\begin{align}
\begin{aligned}
E_{\rm pot}=\frac{1}{2} \int_{\mathcal{V}}{\Phi \rho \mathrm{d}V'} \, ,
\end{aligned}
\label{eqn:GPE}
\end{align}
where $\Phi$ is the gravitational potential. The potential is determined by the mass distribution, and changes in that distribution alter the potential energy. Under the constant moment of inertia approximation, the mass distribution is constant, and the potential energy does not change. Even in previous studies that included a change in moment of inertia to calculate kinetic energy \citep[e.g.,][]{Kaula1964}, changes in potential energy were ignored. In reality, the shape of the body and its potential energy both evolve. 

Changes in internal energy are described by the first law of thermodynamics,
\begin{align}
\begin{aligned}
\mathrm{d}E_{\rm int}=\mathrm{d}Q + \mathrm{d}W \, ,
\end{aligned}
\label{eqn:int}
\end{align}
where $Q$ is heat, and $\mathrm{d}W$ is work done. In a closed simple system (i.e., a system that is macroscopically homogeneous, isotropic, large enough so that surface effects can be ignored, not subject to electromagnetic or gravitational forces, and that can exchange energy but not matter with its environment) undergoing quasi-static processes 
\begin{align}
\begin{aligned}
\mathrm{d}E_{\rm int}=T\mathrm{d}S - p\mathrm{d}V \, ,
\end{aligned}
\label{eqn:int_simple}
\end{align}
where $T$ is temperature, $S$ is entropy, and $p$ is pressure. Changes in heat can be driven by: radiation; radioactive decay (although this is comparatively small over the timescales we consider); viscous dissipation (including tidal heating); and the latent heats of phase changes. Although various of these terms in $\mathrm{d}Q$ have been taken into account in previous work on different stages of the recovery after giant impacts \cite[e.g.,][]{Zahnle2015,Lebrun2013,Elkins-Tanton2008,Peale1978}, work done on material in the body has not. However, \citet{Lock2019pressure} showed that changes in shape can lead to substantial changes in pressure (by 10s of GPa at the core-mantle boundary). Compression of material changes its volume and work is done. The work term in Equation~\ref{eqn:int} cannot be ignored during the recovery of bodies after giant impacts.

\section{Methods}
\label{sec:methods}

\subsection{Smoothed particle hydrodynamics}
\label{sec:methods:SPH}

To calculate the energy budget immediately after the impact, we analyzed post-impact bodies produced by the smoothed particle hydrodynamic (SPH) giant impact simulations presented by \citet{Lock2017} and \citet{Lock2019pressure} (Section~\ref{sup:sec:methods:SPH}). These simulations were performed using the GADGET-2 SPH code \citep{Springel2005} modified for planetary impact calculations \citep{Marcusthesis}, which has been used in several studies \citep[e.g.,][]{Cuk2012,Lock2017}. The colliding bodies were differentiated (2/3 rocky mantle, 1/3 iron core by mass) with pure forsterite mantles and pure iron cores modeled using M-ANEOS equations of state \citep{Melosh2007,Canup2012}. See \citet{Cuk2012} for a full description of methods. These collisions cover a large range of impact parameters that produce approximately Earth-mass bodies. The resulting structures have a wide range of thermodynamic and rotational states (Table~\ref{sup:tab:impacts}).

To quantify the energy of each impact, we use an empirical modified specific energy, $Q_{\rm S}$ \citep{Lock2017}. $Q_{\rm S}$, is defined by 
\begin{linenomath*}
\begin{equation}
Q_{\rm S} = Q'_{\rm R} \left ( 1 + \frac{M_{\rm p}}{M_{\rm t}} \right ) (1-b) \,,
\label{eqn:QS}
\end{equation}
\end{linenomath*}
where $Q'_{\rm R}$ is a center-of-mass specific impact energy modified to include only the interacting mass of the projectile \citep{Leinhardt2012}, $M_{\rm p}$ and $M_{\rm t}$ are the mass of the projectile and target respectively, and $b$ is the impact parameter (see Section~\ref{sup:sec:methods:SPH}). Each factor in Equation \ref{eqn:QS} accounts for a facet of how efficiently energy is coupled into the shock pressure field in the impacting bodies. The specific entropy of the mantles of post-impact bodies scales well with $Q_{\rm S}$, making it a good predictor of post-impact thermal state \citep{Lock2017}. 

We qualitatively determined whether each body was above (super-CoRoL) or below (sub-CoRoL) the CoRoL based on the radial angular velocity profile in the midplane in the same manner as in \citet{Lock2017}. In some cases, it was not possible to determine whether the body was above or below the CoRoL and we classified these bodies as co-CoRoL.

\subsection{HERCULES}
\label{sec:methods:HERCULES}

To determine the evolution in the energy budget after an impact, we calculated the properties of corotating planets with varying thermal and rotational states (Section~\ref{sup:sec:methods:HERCULES}) using the HERCULES code \citep{Lock2017}. HERCULES uses a potential field method to calculate the axisymmetric equilibrium structure of planets with a given thermal state, composition, mass and AM, using realistic equations of state. \citet{Lock2017} found no evidence for triaxial Earth-like bodies even at very high AM, and so the assumption of axisymmetric bodies is likely valid.

To allow direct comparison to the SPH post-impact bodies, we used the same equations of state as for the impact simulations. The energies of corotating bodies calculated using SPH and HERCULES agree to within a few percent (Section~\ref{sup:sec:methods:comparison}). We considered bodies with two different thermal profiles: one for partially-vaporized bodies and another for condensed planets (Section~\ref{sup:sec:methods:HERCULES}). To calculate the properties of partially-vaporized bodies at the CoRoL (stage II), we imposed thermally stratified mantles that emulate the thermal structure of bodies after an impact and during condensation \citep{Lock2017,Lock2018moon}. The core was assumed to be on an isentrope similar to the present-day. For a body of a given AM, mass, and core-mass fraction, the thermal profile at which the body is at the CoRoL was found by determining when the angular velocity of the equator is equal to a Keplerian orbit by linearly extrapolating from an array of calculated planets (Section~\ref{sup:sec:methods:HERCULES}). The properties of the planet at the CoRoL (i.e., the planet with the determined maximal thermal profile) were then found by extrapolation from the same array of calculated planets to the maximal thermal profile.

For condensed bodies (stages III to V), we used a mantle isentrope that intersects the liquid-vapor phase boundary at 10 bar and 4000~K. This thermal state is that of a well-mixed, liquid, magma-ocean planet with a thin volatile-dominated atmosphere, similar to that used in previous work \citep{Elkins-Tanton2008,Lebrun2013}. The volume change upon freezing of silicates is small, particularly at high pressures, and freezing the mantle has little effect on the kinetic and potential energy of the body (Section~\ref{sup:sec:methods:HERCULES}). As we do not consider the change in internal energy upon crystallization of the mantle (transition from stages III to IV), we use the same thermal structure for all condensed bodies. Therefore, our reference Earth model will be a corotating magma-ocean planet of Earth mass and composition throughout.

\section{Results}
\label{sec:results}

\subsection{Cooling of the post-impact body}
\label{sec:results:cooling}

\begin{figure*}
\centering
\includegraphics[height=0.95\textheight]{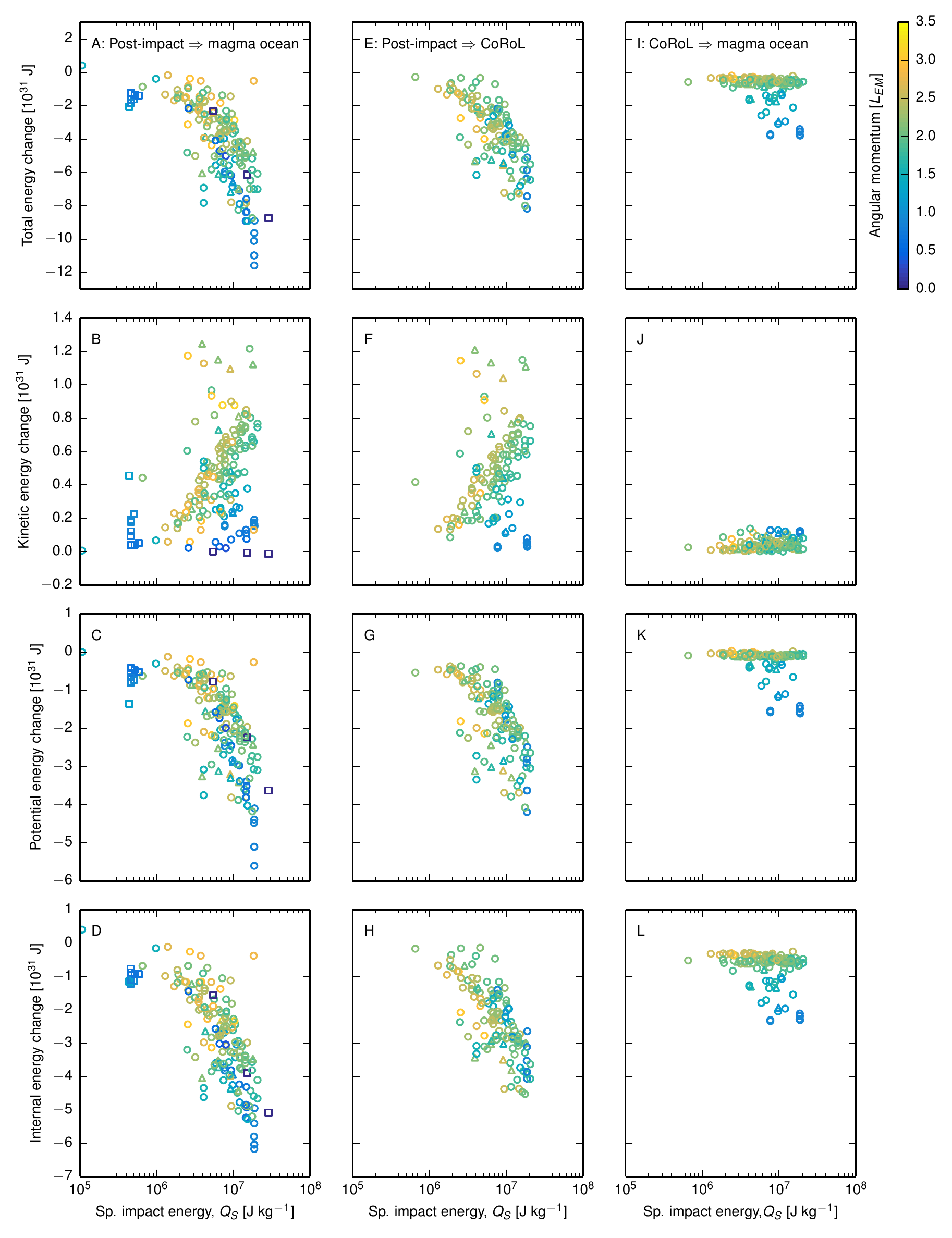}
\caption{Caption on next page.}
\end{figure*}
\setcounter{figure}{1}
\begin{figure*}
\centering
\caption{Condensing the silicate vapor required the loss of substantial amounts of energy. If the post-impact body was initially super-CoRoL, the majority of this energy must have been lost for the body to fall below the corotation limit (CoRoL). Left column: the change in the total energy (A) and energy components (B-D) due to the condensation of post-impact bodies. Middle column: the energy change in cooling initially super-CoRoL bodies to the CoRoL. Right column: the energy change in cooling initially super-CoRoL bodies from the CoRoL to a magma-ocean planet. For reference, the total, potential, kinetic, and internal energies of post-impact bodies are typically on the order of negative $10^{32}$, negative $10^{32}$, $10^{31}$ and $10^{31}$~J respectively. Post-impact bodies with a final bound mass between $0.9$ and $1.1 M_{\rm Earth}$ are plotted as a function of the geometrically-modified specific energy, $Q_{\rm S}$, of the impact that produced them (Equation~\ref{eqn:QS}). Colors indicate the angular momentum of the post-impact bound mass, in units of the present-day angular momentum of the Earth-Moon system ($L_{\rm EM}$). Symbols show structures that are above ($\circ$), below ($\Box$), or have an unclear relationship to ($\bigtriangleup$) the CoRoL \citep{Lock2017}. Bodies that were not initially above the CoRoL are not plotted in the center and right columns, but we have included those cases identified as co-CoRoL.}
\label{fig:Echange_cooling}
\end{figure*}

Post-impact bodies must radiate a considerable amount of energy in order to condense. The first column of Figure~\ref{fig:Echange_cooling} shows the energy difference between bodies immediately after collision and the corresponding magma-ocean planets (i.e., the change in energy going from stage I to III). We ignored the effect of satellite formation as it is uncertain and comparatively small (Section~\ref{sup:sec:Mooneffect}). Energy changes are plotted against the modified specific energy of the impact that produced each body (Equation~\ref{eqn:QS}). Colors indicate the AM of the bound mass. For comparison, the total, potential, kinetic, and internal energies of post-impact bodies are typically on the order of negative $10^{32}$, negative $10^{32}$, $10^{31}$ and $10^{31}$~J respectively, and the heat that has been produced in Earth by radioactive decay since its formation is on the order of several $10^{30}$~J. 

The energy change upon condensation varies substantially between different impacts, with fractional changes in the total energy budget from one to over a hundred percent. Note that fractional decreases of over 100\% are possible as the gravitational potential energy is defined as negative. The changes in potential (Figure~\ref{fig:Echange_cooling}C) and internal (Figure~\ref{fig:Echange_cooling}D) energy typically dominate. A large mass fraction of post-impact bodies can have specific entropies above the critical point, up to half the silicate mass \citep{Lock2018moon}, and the internal energy that must be lost to cool the structure to a liquid isentrope is substantial, on the order of tens of percent of the initial internal energy. The pressure increase upon cooling \citep{Lock2019pressure} does compressive work which partly offsets the decrease in heat, but this term is small compared to the latent heat of vaporization. Potential energy is released as post-impact bodies cool and contract as more mass is focused at the center of the body and the gravitational well deepens (Figure~\ref{fig:pcontours}A-C). The fractional change in gravitational potential energy can be as high as 30\% but is typically on the order of a few percent. Contraction reduces the moment of inertia and, in order to conserve AM, the rotation rate of the magma-ocean planet is higher than the average rotation rate of material in the post-impact body \citep{Lock2019pressure}. A quickening of rotation increases the kinetic energy. This effect is greatest for high AM bodies (Figure~\ref{fig:Echange_cooling}B). The kinetic energy increase is typically several times smaller than other terms, except for high-AM bodies produced by low-energy impacts. 

Because the internal and potential energy terms are dominant, the energy change upon condensation is largely controlled by the specific energy of the impact. High-energy collisions produce hotter bodies that tend to be extended, with a substantial mass fraction supported by thermal pressure far from the center of mass. The changes in internal and potential energy are greater than for colder, more compact structures produced by lower-energy impacts. The energy budget is hence substantially different for the various Moon-formation models. Intuitively, after a modest-energy canonical impact less energy loss is required to condense than in proposed high-AM scenarios.

For bodies that are initially synestias, most of the condensation and corresponding change in shape occurs before the body cools to below the CoRoL. The middle column of Figure~\ref{fig:Echange_cooling} shows the energy difference between immediately after the impact (stage~I) and when the body has cooled to the CoRoL (stage II). Continued cooling to a magma ocean (stage III) typically causes only a small energy change (Figure~\ref{fig:Echange_cooling}, right column), except for bodies with lower AM ($L\lesssim 1.5L_{\rm EM}$). In such cases, the CoRoL is crossed when the mantle is still substantially vaporized and more cooling is required after crossing the CoRoL to reach a fully-liquid state.

Using our calculations, we estimated a lower limit for the timescale of condensation. The time required to condense a post-impact body is dependent on the radiating surface area ($A$), its radiative temperature ($T_{\rm rad}$), and the effective emissivity ($\epsilon$). The photospheric temperature immediately after the impact is about 2300~K \citep{Lock2018moon}. As the body cooled, the outer regions became increasingly volatile dominated and the photospheric temperature dropped substantially \citep{Lupu2014}. Nevertheless, the transition from silicate to volatile-dominated occurred at thermal states close to that which we prescribed for our magma-ocean planets, and the radiative temperature likely remained high for most of condensation. We use a photospheric temperature of 2300~K to provide a lower limit on the cooling timescale. Radiative heat loss during condensation was likely dominated by molten silicate droplets \citep{Lock2018moon} which have emissivities close to unity \citep{Root2018}, but the effective emissivity could be lowered by the presence of dust above the photosphere.

\begin{figure}
\centering
\includegraphics{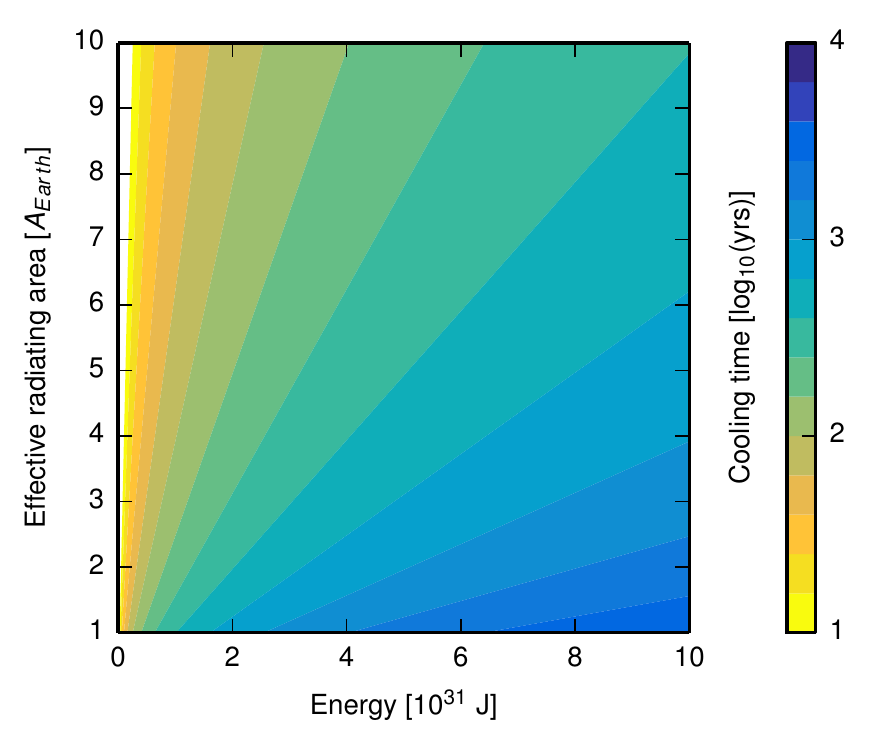}
\caption{A minimum of hundreds to thousands of years is required to radiate sufficient energy to condense a post-impact body. Shown are contours of the time required to radiate an amount of energy given an effective surface area (the true surface area, $A$, multiplied by the effective emissivity, $\epsilon$) and a radiative temperature of 2300~K. \citet{Lock2018moon} showed that the radiative temperature of a vapor-dominated silicate atmosphere is controlled by radiation from silicate droplets at about 2300~K. The effect of solar radiation is negligible and so was not included in the calculation.
}
\label{fig:time_conversion}
\end{figure}

Figure~\ref{fig:time_conversion} shows the time taken to lose an amount of energy by radiation given an effective radiative area, $\epsilon A$. The surface area immediately after a collision can be many times that of the present-day Earth, $A_{\rm Earth}$. The outer regions of post-impact bodies typically contain only a small fraction of the mass and condense quickly, resulting in rapid contraction of the body \citep{Lock2018moon}. The cooling timescale is dominated by the later evolution when the radiative surface area was likely a few times that of the present-day Earth, but dynamical simulations are required to confirm this supposition. Based on this assumption, most post-impact bodies will take at least 100s to 1000s of years to condense. For bodies that are initially synestias, most of that time is taken cooling to the CoRoL.

\subsection{Lunar tidal recession}
\label{sec:results:tidal}

Following the formation of the Moon, tidal torques between Earth and the Moon forced the lunar orbit to evolve and over time pushed the Moon farther away from Earth. In order to compensate for the increased AM of the lunar orbit and due to other orbital interactions, the rotation rate of Earth was reduced. Tidal evolution increased the energy of the lunar orbit but decreased the total energy of the Earth. For the lunar orbit to evolve, energy must have been dissipated as heat in Earth and the Moon. The rate of tidal evolution was therefore largely dependent on how quickly energy could be dissipated in the two bodies. In order to determine the rate of tidal recession and the effect of tidal evolution on the thermal structure of Earth and the Moon, it is necessary to determine the energy budget of the system and how it changes as the lunar orbit evolves.

Previous work \citep[e.g.,][]{Zahnle2015,Touma1994} only considered changes in the Moon's orbital energy, Earth's kinetic energy, and the heat component of internal energy (Equation~\ref{eqn:int}). However, the change in Earth's figure as it slowed also altered its potential energy, and the internal pressure increase did $p\mathrm{d}V$ work. Here, we focus on understanding the effect of changes in shape on the potential, kinetic and internal energy of Earth during tidal recession. In calculating the internal energy, we ignored changes in heat (Equation~\ref{eqn:int}), e.g., due to the dissipation of tidal energy, to consider only the $p\mathrm{d}V$ work done by increasing pressure. 

\begin{figure}
\centering
\includegraphics{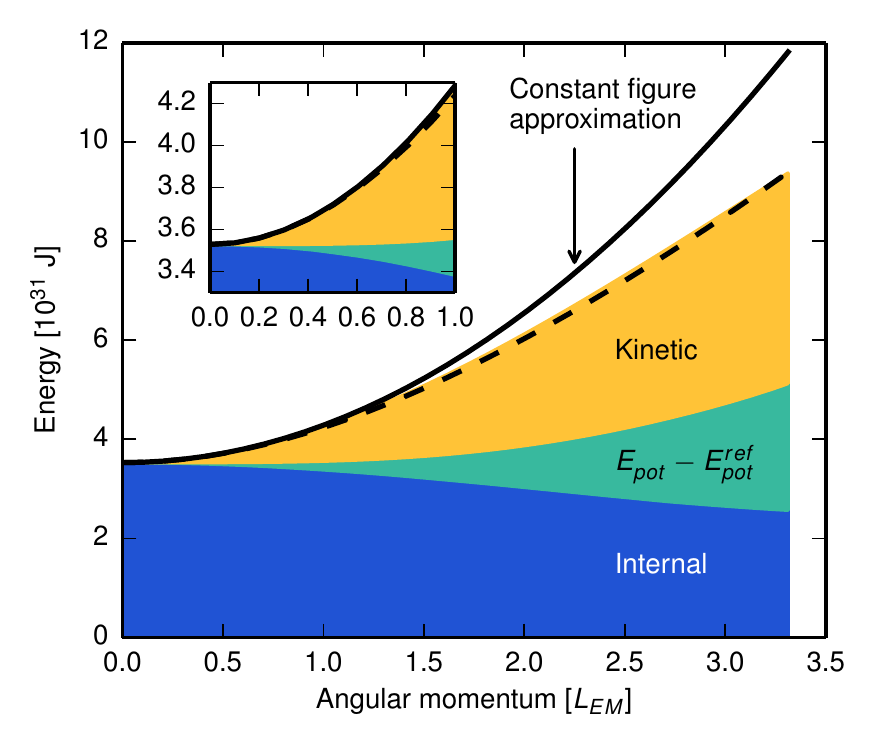}
\caption{Angular momentum transfer during lunar tidal recession changed the kinetic, potential, and internal energies of Earth. Colored areas show the energy components of an Earth-like magma-ocean planet as a function of angular momentum. The potential energy is referenced to that of a non-rotating body. The black lines show the energy using the commonly made assumption that the potential and internal energies of Earth remained constant and a constant moment of inertia (solid) or small-deformation approximation (dashed, Equation~\ref{eqn:MOI}). The inset focuses on bodies with modest initial angular momenta, such as in the canonical Moon-formation scenario. Note that the internal energy in this figure is calculated using an ANEOS equation of state that only includes a simple high-pressure phase transition \citep{thompson1972}, and we have not included any heat loss or gain mechanisms, i.e., the heat term ($\mathrm{d}Q$) in Equation~\ref{eqn:int} was assumed to be zero.
}
\label{fig:Echange_despin}
\end{figure}

Figure~\ref{fig:Echange_despin} shows the energy budget of corotating Earth-like magma-ocean planets with different AM. For clarity, the potential energy is shown relative to that of a non-rotating body. For comparison, the orbital energy released due to changes in the Moon's orbit is about $8 \times 10^{29}$~J since its formation. Earth today has an AM of 0.18~$L_{\rm EM}$, and when the Moon was at the Cassini-state transition \citep[which occurred when the lunar semi-major axis was about 30 Earth radii,][]{Ward1975} Earth's AM was 0.417~$L_{\rm EM}$. The solid black line shows the result making the assumption that the moment of inertia of Earth is constant and only the kinetic energy is changing. The black dashed line shows another approximation that only includes kinetic energy but allows for evolution in Earth's shape. The Earth's moment of inertia is calculated, assuming small deformation, as
\begin{equation}
I=I_{\omega=0}+\left ( \frac{2k_{\rm Earth}R_{\rm Earth}^5}{9G} \right ) \omega^2 \, ,
\label{eqn:MOI}
\end{equation}
where $I_{\omega=0}$ is the moment of inertia of a non-rotating body, $k_{\rm Earth}=0.94$ is the secular Love number of Earth \citep{Yoder1995}, and $R_{\rm Earth}$ is the radius of Earth \citep[e.g.,][]{Munk1975}.

\begin{figure}
\centering
\includegraphics{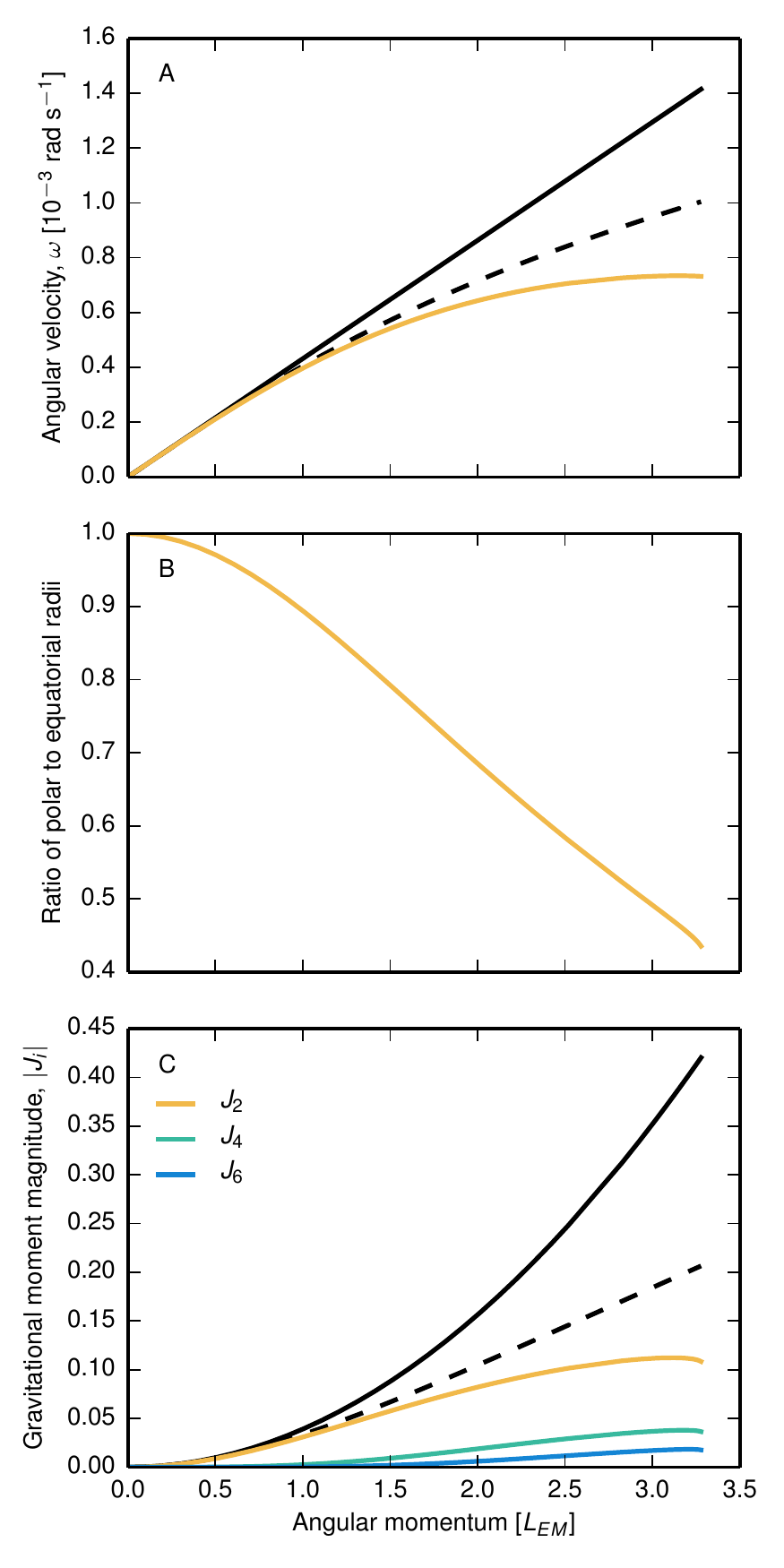}
\caption{The shape of Earth changed significantly with angular momentum, altering its energy budget and changing the relation between angular momentum, rotation rate and other physical parameters. The relationship between angular momentum and the angular velocity (A), ratio of polar to equatorial radii (B), and the first three even gravitational moments (C) for an Earth-like magma-ocean planet are presented. For comparison, the solid black line shows the properties calculated assuming Earth had a constant moment of inertia equal to that of a non-rotating body, and the dashed black line shows the properties calculated assuming that the moment of inertia of Earth changed as described by Equation~\ref{eqn:MOI}. In both cases, it was assumed that $J_2 \sim \omega^2$ and all other gravitational moments were zero.
}
\label{fig:figure_change_despin}
\end{figure}

Bodies with high AM have more energy than slowly rotating bodies, and, as the rotation rate of Earth slowed, energy must have been dissipated as heat in Earth and the Moon. The kinetic energy component changed the most, albeit less than has been calculated using constant-moment of inertia and small deformation assumptions. At higher AM, bodies are substantially oblate (Figure~\ref{fig:figure_change_despin}B) and the moment of inertia is larger. The rotation rate at a given AM is correspondingly slower (Figure~\ref{fig:figure_change_despin}A) and the kinetic energy is lower (Equations~\ref{eqn:KE_corot_omg} and \ref{eqn:KE_corot_L}). As the rotation rate decreases, Earth's shape changes and potential energy is released. Work done by adiabatic compression raises the internal energy, partially offsetting changes in other energy terms. The equation of states used in our calculations \citep[M-ANEOS forsterite and iron,][]{Melosh2007,Canup2012} include the liquid-vapor phase boundary (which is not relevant for the magma-ocean planets we consider here) and a simple high-pressure phase transition \citep{thompson1972} which occurs around 80~GPa in the mantle. Additional pressure-induced phase transitions would lead to more substantial increases in internal energy (see Section~\ref{sec:disc:energy_disp}). It is only fortuitous that the change in kinetic energy calculated using the small deformation formulation (Equation~\ref{eqn:MOI}) gives a good approximation to the change in the total energy. 

After formation of the Moon following a canonical giant-impact, Earth had a modest AM of $\lesssim 0.8 L_{\rm EM}$. The largest change during lunar tidal recession would have been in kinetic energy, but there would also have been changes in potential and internal energy of about 20\% of the magnitude of the change in kinetic energy (inset in Figure~\ref{fig:Echange_despin}). After high-AM Moon forming impacts, the larger change in figure during tidal recession results in greater changes in potential and internal energy. The total change in energy is a few times larger than in the canonical case. In all scenarios, the energy released during tidal recession would have been comparable to, or larger than, the energy required to melt Earth's mantle ($0.16 \times 10^{31}$~J, assuming a latent heat of $4 \times 10^{5}$~J~kg$^{-1}$). Current models predict that exchange of AM occurred mostly in the first couple of 10s~Myr \citep{Touma1994,Zahnle2015,Cuk2012,Wisdom2015,Cuk2016,Tian2017}. For comparison, the energy released by radioactive decay in Earth during that time was on the order of $10^{28}$~J. 

\section{Discussion}
\label{sec:disc}

\subsection{Condensation of post-impact bodies}
\label{sec:disc:cooling}

The stage of evolution immediately after the Moon-forming impact, condensation of the silicate vapor, has not been considered in detail. Typically, cooling calculations begin when Earth was a magma-ocean planet with a volatile-dominated atmosphere \citep[e.g.,][]{Elkins-Tanton2008,Lebrun2013}. Studies that approximated the time taken to condense have done so based on a model of purely latent heat extraction by radiation, assuming a spherical body the size of the present-day Earth \citep[e.g.,][]{Zahnle2015}. No other processes that change the energy budget have been accounted for.

We have quantified the energy that needs to be lost in order for a post-impact body to condense, including changes in shape, and pressure and temperature structure. The change in energy is substantial, an order of magnitude greater than radioactive heat production over the lifetime of Earth, and changes in kinetic, potential, and internal energy can all be significant. Only considering changes in internal energy underestimates the total energy change by up to a factor of two. 

Changes in kinetic and potential energy alter how the thermal profile of the post-impact body evolved. Potential energy would have been dissipated heterogeneously throughout the body as it contracted and the potential well deepened. Shear would have propagated the change in rotation rate across the body, again dissipating energy heterogeneously. Here we have only considered stages in evolution, and a dynamical calculation will be required to determine the effect of these extra energy terms on Earth's thermal profile. Understanding this process is particularly important to assess the survival of distinct geochemical signatures that predate Moon formation \citep{Rizo2016a,Mukhopadhyay2012,Peto2013,Tucker2012,Parai2012,Mundl2017}. 

Based on our calculations, Earth would have remained substantially vaporized for at least 100s to 1000s years after the Moon-forming impact. This is on the same order as the simplified estimates given in previous work \citep[e.g.,][]{Lock2017}. Condensation after a high-energy, high-AM impact takes longer than in the canonical scenario. The synestias formed in such impacts would remain above the CoRoL for the majority of condensation. Although the period of condensation was short, it came at a key point in Earth's evolution and the aberrant structure of the body would have influenced several processes including Moon formation, reaccretion of debris, and atmospheric loss. 

The body would have been substantially vaporized and spatially extended throughout the period of Moon formation \citep[10s to 100s years, e.g.,][]{Salmon2012,Lock2018moon}. \citet{Lock2018moon} showed a substantially-vaporized synestia provides a new environment for lunar formation, but the presence of large amounts of vapor would also have affected Moon formation in the canonical scenario. The inner boundary of the lunar disk would not have been a condensed body about the size of the present-day Earth, as is often assumed in disk models \citep[e.g.,][]{Ida1997,Kokubo2000,Salmon2012}. The large radius of the corotating region of the structure would have imposed an inner edge to the disk farther away from the center of mass and closer to the Roche limit, the radius beyond which a satellite can form \citep{Lock2017}. A significant vapor fraction would also have altered the disk dynamics \citep{Desch2013,Lock2017,Lock2018moon}. Future work must consider the effect of the long-lived hot thermal state of the post-impact body on Moon formation from a canonical disk.

The collisional cross sections of bodies immediately after an impact can be tens of times larger than slowly-rotating, condensed planets \citep{Lock2017}. After the Moon-forming impact, the probability of impacts would have been increased while the body was still partially vaporized. Furthermore, the collisional cross section of highly oblate, condensed planets can also be up to a factor of two higher than slowly-rotating planets, and so the impact flux onto Earth would have remained elevated during the early stages of lunar tidal evolution after a high-AM Moon-forming impact. Using the present-day cross section of Earth and a simple model for ejecta, \cite{Jackson2012} found that 0.5\% of debris from the impact would be accreted within a 1000~yrs, i.e., within the timescale of condensation. Given the substantially larger cross section of the post-impact Earth and that up to $\sim 10$\% of an Earth mass of ejecta can be produced by giant impacts \citep{Canup2001,Canup2012,Cuk2012}, a significant mass of impact debris could have been accreted while Earth was still substantially vaporized. Earth may have also accreted a significant mass of other material leftover from the main stage of planet formation as the impactor flux was likely high during this time.

In the vapor regions of the body, metal was soluble in the silicate vapor \citep{Lock2018moon}. Any metal delivered by impactors into the substantially vaporized outer regions of the post-impact Earth would be, at least temporarily, incorporated into the bulk silicate Earth (BSE). Some metal could exsolve and be segregated to the core during cooling, but determining the efficiency of this process requires modeling the dynamical and chemical evolution of the body. Conversely, metal injected to higher-pressure in the partially vaporized structure would likely be quickly incorporated into the core, as the residence time for iron droplets in a fluid mantle \citep[$\leq 100$~yrs,][]{Stevenson1990} is shorter than the cooling timescale. Iron accreted by Earth after condensation of the vapor but before the lower mantle solidified would also likely be incorporated into the core. If significant mass were accreted during condensation, these processes would complicate the interpretation of the terrestrial budget of highly siderophile elements as a limit on the mass accreted since lunar formation \citep[see review by][]{Morbidelli2014}. Furthermore, projectiles hitting Earth before the silicate vapor had condensed would have been less efficient at removing volatiles. Calculations of the efficiency of atmospheric loss assume that the body has a condensed surface where most of the energy is deposited, driving shocks through the planet and atmosphere \citep{Schlichting2012,Schlichting2018}. Impacts into a partially-vaporized body would be slowed gradually by the silicate vapor and been less effective in removing volatile elements, aiding in the retention of volatiles.

\subsection{Energy dissipation during lunar tidal recession}
\label{sec:disc:energy_disp}

\begin{figure}
\centering
\includegraphics[]{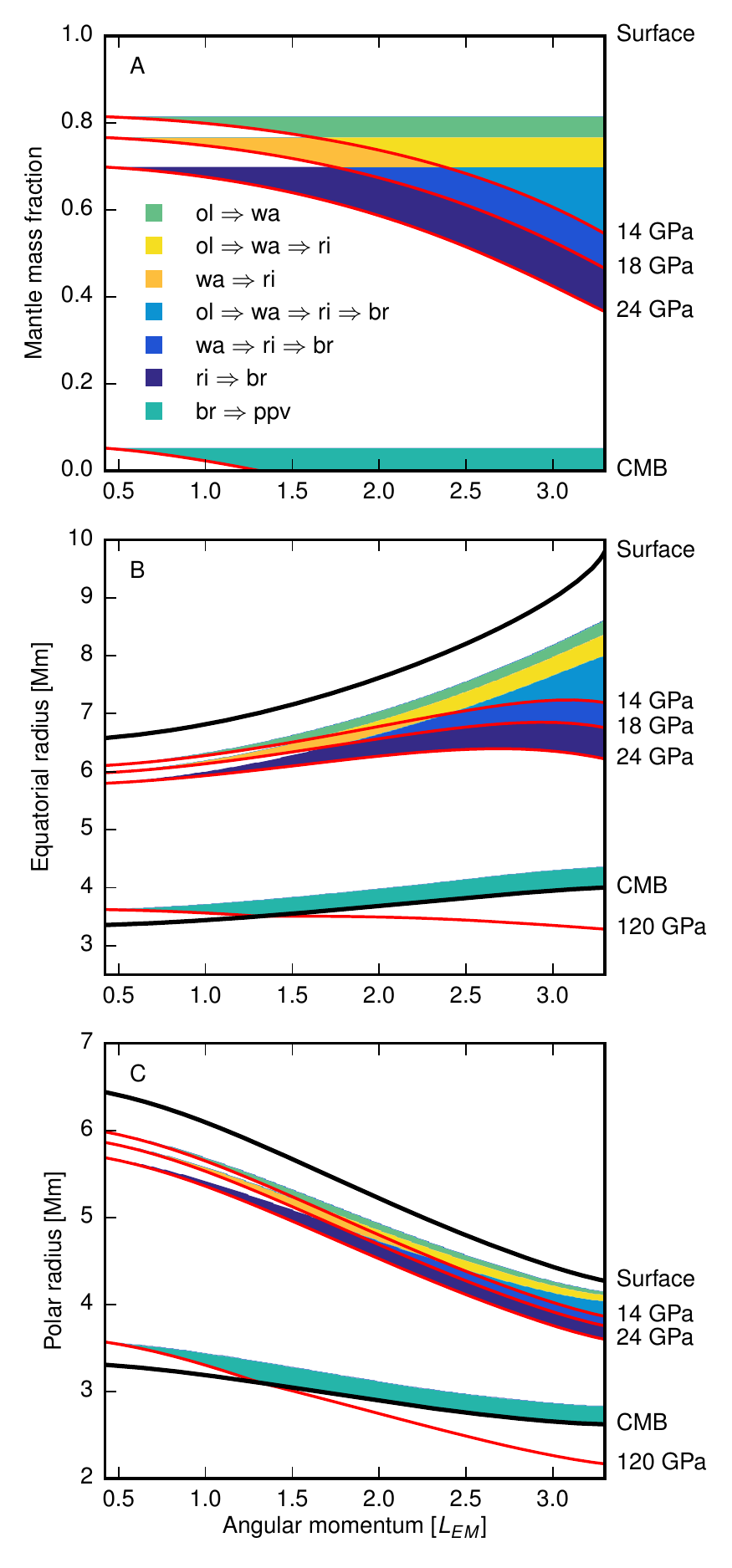}
\caption{Caption opposite.}
\end{figure}
\setcounter{figure}{5}
\begin{figure}
\centering
\caption{Increases in pressure during tidal recession induce phase changes, concentrating the rise in internal energy in specific regions of the mantle. (A) The mass-fractions of the mantle of a body, starting at a given angular momentum, that undergo phase transitions as the angular momentum of the body decreases to 0.417~$L_{\rm EM}$, the approximate angular momentum of Earth when the lunar spin axis underwent the Cassini-state transition \citep[which occurred when the lunar semi-major axis was about 30 Earth radii,][]{Ward1975}. Mantle mass fraction is defined as zero at the core-mantle boundary and one at the surface. Colors show the portions of the mantle that cross the pressures of phase transitions between olivine (ol), wadsleyite (wa), ringwoodite (ri), bridgmanite (br), and post-perovskite (ppv). Large portions of the mantle could have undergone multiple phase transitions. For the purposes of this illustration, we have assumed that the phase transitions occur at similar pressures as they occur in the present-day mantle. Red lines show the mass fraction that is above the pressure of each phase transition (see text for details). The lower panels show the same information but referenced to the radius in the equatorial plane (B) and at the poles (C). Black lines show the radius of the core-mantle boundary and surface.
}
\label{fig:phase_transitions}
\end{figure}

During lunar tidal recession, energy was dissipated as heat in both Earth and the Moon \citep[e.g.,][]{Peale1978}. The energy released was substantial (comparable to that required to melt the mantle) and significantly perturbed the thermal structure of both bodies. Energy was dissipated heterogeneously with potentially long-lasting effects on the thermal evolution of Earth. There is also a feedback between the thermal state of a body and its tidal parameters \citep[e.g.,][]{Peale1978,Zahnle2015,Henning2014} which could have controlled the rate of lunar tidal recession and hence energy dissipation. Quantifying the magnitude and mechanism of energy dissipation during tidal recession is key to understanding the evolution of early Earth.

Previous work only considered changes in the orbital energy of the Moon and the kinetic energy of Earth, which evolve due to torques between the Sun, Moon and Earth. Under these assumptions, the heat dissipated in a portion of Earth is dependent on its rheological properties and position. However, we have shown that changes in shape substantially altered Earth's energy budget during lunar recession. Accounting for the changing shape, the energy released due to a given change in AM is less than has typically been assumed. The difference would have been largest if the Earth-Moon system started with higher AM than the present day. There were also substantial changes in potential and internal energy as well as kinetic energy.

Changes in energy components other than kinetic energy altered how the thermal structure of Earth evolved. Potential energy was released as material fell further into the gravity well as the centrifugal force diminished (green region in Figure~\ref{fig:Echange_despin}). The potential energy was not altered by work done by tidal torques, but rather released as a by-product of the change in rotation driven by these torques. The potential energy released due to the changing shape of Earth  was deposited only in Earth, over the whole body, in contrast to orbital and kinetic energy which were dissipated in both Earth and the Moon.

Increases in internal energy due to $p\mathrm{d}V$ work offset some of the decrease in other energy components. We calculated the internal energy change due to compression of a single-phase material, using equations of state that only include vaporization and a simple high-pressure phase transition \citep{thompson1972} in the mantle and do not include a full treatment of other phase transitions. However, \citet{Lock2019pressure} suggested that pressure increases during tidal recession could cause both melting and freezing of different sections of the mantle. Increases in pressure would also have driven solid-solid phase transitions over a large fraction of Earth's mantle. BSE-composition material undergoes a number of exothermic and endothermic phase transitions between about 14 and 25~GPa (in the transition zone between 410 and 660~km depth in the mantle today) and there is a transition of bridgmanite to post-perovskite at about 120~GPa \citep[e.g.,][]{Stixrude2011}. Figure~\ref{fig:phase_transitions} illustrates the fraction of the mantle that could have undergone phase transitions during tidal recession from a given initial AM to the Cassini-state transition. Phase boundaries are dependent on temperature and, as the thermal profile during tidal recession is uncertain, for this illustrative calculation we used phase-transition pressures similar to those in the present-day mantle. We considered: the olivine $\Rightarrow$ wadsleyite transition at 14~GPa; the wadsleyite $\Rightarrow$ ringwoodite transition at 18~GPa; the ringwoodite $\Rightarrow$ bridgmanite $+$ periclase transition at 24~GPa; and the bridgmanite $\Rightarrow$ post-perovskite transition at 120~GPa. Note that majorite also transforms to bridgmanite at about 24~GPa. For initially high-AM bodies, tens of percent of the mantle would have undergone phase transition during tidal recession. Parts of the mantle would have changed phase multiple times.

For adiabatic compression, $\mathrm{d}Q=0$ and Equation~\ref{eqn:int} reduces to
\begin{equation}
\mathrm{d}E_{\rm int}=-p\mathrm{d}V \, .
\end{equation}
Pressure-driven phase changes typically decrease the specific volume of material and would have increased Earth's internal energy. Additional phase transitions would have added to the internal energy gain that we calculated and further reduced the amount of energy dissipated as heat due to decreases in other energy components. Internal energy increases would have been focused in the areas of the mantle which underwent phase changes (Figure~\ref{fig:phase_transitions}), reducing the net heating in these regions.

Internal energy changes would have influenced not only the magnitude, but also the rate and spatial distribution of tidal heating. For example, it has been argued that tidal dissipation would have been focused in partially-molten regions of the mantle due to their intermediate viscosity, setting up a feedback between melting and tidal dissipation \citep[e.g.,][]{Zahnle2015}. However, increases in pressure during tidal recession may have damped this feedback. Compressive work would have been done in partially molten regions as the pressure increased, reducing the energy available to increase the melt fraction. In other words, pressure-induced freezing would have acted against tidal heating to reduce the melt fraction of partially-molten regions. 

Determining the mechanisms by which changes in kinetic, potential and internal energy were enacted, and how these mechanisms are related, is necessary for understanding the thermal evolution of Earth during lunar tidal recession.

\subsection{Influence of a changing figure on orbital dynamics}
\label{sec:disc:dynamics}

The dynamics of lunar tidal evolution controlled the rate of energy dissipation in Earth and the Moon and how the AM of the system changed. Determining by how much the AM could have changed is vital to constrain the feasible range of Moon-forming impacts. However, a number of approximations are made in calculating tidal evolution. We now consider the effect that the changing figure of Earth has on the validity of these approximations and its consequences for tidal evolution.
 
The tidal quality factors, $Q_{\rm orb}$, of Earth and the Moon dictated the rate of tidal dissipation and hence how quickly the lunar orbit evolved. The tidal quality factors are uncertain due to difficulties in determining the frequency-dependent response of bodies with arbitrary thermal profiles \citep[e.g.,][]{Henning2014}, and complications in simultaneously modeling thermal and tidal evolution \citep{Zahnle2015,Tian2017}. Due to these difficulties, it is often assumed that $Q_{\rm orb}$ and the tidal Love numbers, $k_i$, are constant. The tidal properties of bodies with varying thermal profiles have been calculated by considering perturbations to a spherical body \citep[e.g.,][]{Henning2014,Tian2017}. However, the effective potential field of Earth would have been altered due to the large oblateness of the body and the centrifugal force, particularly after a high-AM impact. For example, the effective gravity at the equator of Earth-like bodies with AM of between $1$ and $3$~$L_{\rm EM}$ ranges from $8$ to $2$~m~s$^{-2}$, much lower than for present-day Earth. The body would have experienced altered tidal stresses and the tidal response for a given rheological distribution could have been substantially different. The Moon would likely have raised a larger tidal bulge on a rapidly-rotating Earth, due to the lower equatorial gravity. A larger tidal bulge with the same degree of lag behind the lunar orbit would have produced stronger tidal torques and accelerated lunar recession. The effect of rotation on the tidal love numbers of the giant planets has already been investigated by \cite{Wahl2017a}, who found a significant increase in $k_2$ with rotation rate, but the effect of rotation after high-AM Moon-forming impacts would be even more extreme. Quantifying the effect of large oblateness on the tidal response of early Earth will require more advanced techniques \citep[e.g.,][]{Al-Attar2018}.

In tidal evolution models to date, Earth's moment of inertia is assumed to be constant \citep[e.g.,][]{Cuk2012,Wisdom2015,Cuk2016,Tian2017} or is estimated based on small deformations from a sphere \citep[e.g.,][]{Kaula1964}. The assumption used determines the relationship between AM, $L$, and $\omega$. At modest AM ($\lesssim 1~L_{\rm EM}$) the assumption of constant moment of inertia approximates well the $L$-$\omega$ relation, but at higher-AM ($\gtrsim 1~L_{\rm EM}$) the rotation rate is significantly slower than for a body with constant moment of inertia (Figure~\ref{fig:figure_change_despin}A). $L$-$\omega$ relationships based on small deformations from a sphere (dashed line in Figure~\ref{fig:figure_change_despin}) are in better agreement, but still diverge at high AM. The effect of a realistic $L$-$\omega$ relation depends on the tidal model used. For models that assume a constant time lag between the tidal bulge and the lunar orbit \citep[e.g.,][]{Touma1994}, $Q_{\rm orb} \propto \omega^{-1}$ and a lower $\omega$ at a given AM would result in a higher $Q_{\rm orb}$ and more rapid tidal recession. For models \citep[such as][]{Cuk2012,Wisdom2015,Cuk2016} that use the constant-$Q_{\rm orb}$ model \citep{Kaula1964,Goldreich1966} the relationship between $\omega$ and the tidal torque is more complicated, but smaller $\omega$ at a given AM will also likely lead to faster recession.

Most tidal models consider changes in the degree-2 gravitational moment as a simple function of rotation rate, $J_2 \propto \omega^2$. This relationship holds well even at high AM, but, due to the altered $L$-$\omega$ relation, the $J_2$ of Earth calculated using constant moment of inertia and small perturbation assumptions is erroneously high. A difference in the $L$-$J_2$ relation changes the dynamical behavior of the Earth-Moon system. For example, the Moon may have entered into an evection resonance \citep{Cuk2012} or a near-evection limit cycle \citep{Wisdom2015} during its orbital evolution, which could have transferred AM away from the Earth-Moon system. The Hamiltonian of the evection resonance contains terms linearly dependent on $J_2$ and others inversely dependent on $L^2$ \citep{Touma1998}. Altering the $L$-$J_2$ relation would have altered the behavior of evection and its ability to transfer AM. An alternative mechanism for transferring AM is an instability during the Laplace-plane transition (where the lunar orbit moves from being dominated by Earth's equatorial bulge to being dominated by solar perturbations) for a planet with initially high obliquity \citep{Cuk2016}. The semi-major axis at which this transition occurs is dependent on $J_2^{1/5}$ \citep{Nicholson2008}, and a different $L$-$J_2$ relation also changes the tidal evolution of an initially high-obliquity body. In addition, higher order gravitational moments ($J_4$, $J_6$) become comparable in magnitude to $J_2$ at high AM (Figure~\ref{fig:figure_change_despin}) and higher-order tides may have played a non-negligible role in orbital evolution. As the Moon is further from Earth during the Laplace-plane transition the effects of an altered $L$-$J_2$ and higher order gravitational moments are likely smaller than in the case of evection.

Change in Earth's shape could have significantly altered lunar tidal evolution, and hence the rate of energy dissipation in Earth.

\section{Conclusion}
\label{sec:conc}

We have shown that changes in figure and thermal structure significantly altered Earth's energy budget during its recovery after the Moon-forming impact. We have, for the first time, fully quantified the energy that needed to be lost in order for the post-impact body to condense. We have shown that changes in kinetic energy, potential energy, internal energy due to compression, and the shape of Earth were all important in determining Earth's thermal structure and the timescales of evolution. Earth would have remained substantially vaporized for at least hundreds to thousands of years after the impact, changing the environment of lunar formation and allowing for more efficient reaccretion of impact debris. 

We have also quantified the changes in Earth's energy budget due to angular momentum evolution during lunar tidal recession. The competing effects of changes in kinetic, potential and internal energy heterogeneously perturbed Earth's thermal structure in a manner that is yet to be determined. Changes in Earth's figure would also have altered the dynamics of lunar tidal evolution and hence the rate of energy dissipation.

\section*{Acknowledgements}
We would like to thank two anonomous reviewers and the editor for comments that improved the clarity of this manuscipt. This work was supported by NESSF grant NNX13AO67H (SJL), NASA grant NNX15AH54G (STS), DOE-NNSA grant DE-NA0002937 (STS), and NASA grant NNX15AH65G (MC). SJL also gratefully acknowledges support from Harvard University's Earth and Planetary Sciences Department and Caltech's Division of Geological and Planetary Sciences. All data needed to evaluate the conclusions in the paper are present in the paper and/or the Supplementary Materials. The modified version of GADGET-2 and the EOS tables are contained in the supplement of \cite{Cuk2012}. The HERCULES code is included in the supporting information of \cite{Lock2017} and is available through the GitHub repositry:  https://github.com/sjl499/HERCULESv1\_user.

\section*{References}
\bibliographystyle{elsarticle-harv} 
\bibliography{References_manually_corrected}

\clearpage
\clearpage
\renewcommand{\thepage}{S\arabic{page}}  
\renewcommand{\thesection}{S\arabic{section}}   
\renewcommand{\thetable}{S\arabic{table}}   
\renewcommand{\thefigure}{S\arabic{figure}}
\renewcommand{\theequation}{S\arabic{equation}}

\setcounter{page}{1}
\setcounter{section}{0}
\setcounter{figure}{0}
\setcounter{table}{0}
\setcounter{equation}{0}
\setcounter{tnote}{0}
\setcounter{fnote}{0}
\setcounter{footnote}{0}
\setcounter{cnote}{0}
\setcounter{author}{0}
\setcounter{affn}{0}

\renewenvironment{abstract}{\global\setbox\absbox=\vbox\bgroup
  \hsize=\textwidth\def\baselinestretch{1}%
  \noindent\unskip\textbf{Contents}
 \par\medskip\noindent\unskip\ignorespaces}
 {\egroup}

\begin{frontmatter}

\title{Supporting information for ``The energy budget and figure of Earth during recovery from the Moon-forming giant impact"}

\begin{abstract}
\begin{enumerate}
\item Supporting Text S1-S2.
\item Figure S1-S7.
\item Table S1.
\end{enumerate}
\end{abstract}

\end{frontmatter}

\section{Supplementary methods}
\subsection{Smoothed particle hydrodynamics}
\label{sup:sec:methods:SPH}

SPH self-consistently calculates the velocity, gravitational potential, and internal energy using a given equation of state (EOS). It is thus straightforward to calculate the energy budget of post-impact bodies. In SPH, Equation~\ref{eqn:KE_base} can be simplified to 
\begin{equation}
E_{\rm K} = \frac{1}{2} \sum_{i}{r_{xy, i}^2 \omega_i^2 m_i } \, ,
\label{eqn:KE_SPH}
\end{equation}
where $r_{xy, i}$, $\omega_i$ and $\rho_i$ are the cylindrical radius, angular velocity and mass of particle $i$. Potential and internal energies are similarly calculated as a sum over all SPH particles: 
\begin{equation}
E_{\rm pot}=\frac{1}{2} \sum_{i}{\Phi_i m_i} \, ,
\label{eqn:GPE_SPH}
\end{equation}
and
\begin{equation}
E_{\rm int}=\sum_{i}{\epsilon_i m_i} \, ,
\label{eqn:U_SPH}
\end{equation}
where $\Phi_i$ and $\epsilon_i$ are the gravitational potential and specific internal energy of each particle as determined from the distribution of mass and the EOS, respectively.

The modified specific impact energy \citep{Lock2017}, $Q_\mathrm{S}$, takes into account the efficiency with which energy is coupled into the impacting bodies, and is defined by,
\begin{linenomath*}
\begin{equation}
Q_{\rm S} = Q'_{\rm R} \left ( 1 + \frac{M_{\rm p}}{M_{\rm t}} \right ) (1-b) \,,
\label{sup:eqn:QS}
\end{equation}
\end{linenomath*}
where $Q'_{\rm R}$ is a center-of-mass specific impact energy modified to include only the interacting mass of the projectile \citep{Leinhardt2012}, $M_{\rm p}$ and $M_{\rm t}$ are the mass of the projectile and target respectively, and $b$ is the impact parameter. $Q'_{\rm R}$ is given by
\begin{linenomath*}
\begin{equation}
  Q'_{\rm R} = \frac{\mu_{\alpha}}{\mu} Q_{\rm R} \;\;.
  \label{eqn:QR_prime}
\end{equation}
\end{linenomath*}
Note that there is a typographical error in the definition of $Q'_{\rm R}$ in equation 13 of \cite{Leinhardt2012}.
$Q_{\rm R}$ is the unmodified center of mass specific impact energy,
\begin{linenomath*}
\begin{equation}
  Q_{\rm R} = \frac{\mu V_{\rm i}^2}{2 M_{\rm tot}} \;\;.
  \label{eqn:QR}
\end{equation}
\end{linenomath*}
The reduced mass is defined as 
\begin{linenomath*}
\begin{equation}
  \mu = \frac{ M_{\rm p} M_{\rm t}}{ M_{\rm tot}} \;\;,
  \label{eqn:mu}
\end{equation}
\end{linenomath*}
and, to consider only the interacting fraction of the projectile, a modified reduced mass is used, given by
\begin{linenomath*}
\begin{equation}
  \mu_{\alpha} = \frac{ \alpha M_{\rm p} M_{\rm t}}{\alpha M_{\rm p} + M_{\rm t}} \;\;.
  \label{eqn:mu_alpha}
\end{equation}
\end{linenomath*}
$M_{\rm tot}$~$=$~$M_{\rm p}+M_{\rm t}$, $V_{\rm i}$ is the impact
velocity and $\alpha$ is the mass fraction of the projectile that is involved in the collision.  
$\alpha$ is defined as
\begin{linenomath*}
\begin{equation}
 \alpha = \frac{m_{\rm interact}}{M_{\rm p}} = \frac{3 R_{\rm p} l^2 -l^3}{4R_{\rm p}^3}  \;\;,
 \end{equation}
 \end{linenomath*}
where $m_{\rm interact}$ is the interacting projectile mass, $R_{\rm t}$ and $R_{\rm p}$ are the radii of the target and projectile and $B$~$=$~$(R_{\rm t}+R_{\rm p}) b$.   
$l$ is the projected length of the projectile overlapping the target, 
\begin{linenomath*}
\begin{equation}
l = \begin{dcases*}
      R_{\rm t}+R_{\rm p}-B   & when $B+R_{\rm p} > R_{\rm t}$\\
	2R_{\rm p}   & when $B+R_{\rm p} \leq R_{\rm t}$	
        \end{dcases*} \;\;.
\end{equation}
\end{linenomath*}
If $B+R_{\rm p}$~$\leq$~$R_{\rm t}$ then the whole projectile is interacting with the target and $\alpha$~$=$~$1$.

\subsection{HERCULES}
\label{sup:sec:methods:HERCULES}

In HERCULES, a body is described as a series of nested, concentric, constant-density spheroids. All the material between the surfaces of any two consecutive spheroids is called a layer. The energy components are calculated as a sum over these layers. The kinetic energy of a corotating body is simply
\begin{align}
\begin{aligned}
E_{\rm K, co} = & \sum_{i=0}^{N_{\rm lay}-1} \, \frac{1}{2}  I_i \omega_{\rm rot}^2  \, ,
\end{aligned}
\label{eqn:KE_H}
\end{align}
where $N_{\rm lay}$ is the number of spheroids, $I_i$ is the moment of inertia of spheroid $i$, and $\omega_{\rm rot}$ is the angular velocity of the body. Spheroids are numbered from the outside inwards starting at $i=0$.
To calculate the potential energy, Equation~\ref{eqn:GPE} can be approximated as
\begin{align}
\begin{aligned}
E_{\rm pot} = & \sum_{i=0}^{N_{\rm lay}-2} \, \frac{1}{2} \left [ \frac{\Phi_i + \Phi_{i+1}}{2} \right ] \rho_i \left [ V_i - V_{i+1} \right ] \\
& + \frac{1}{2} \left [ \frac{\Phi_{N_{\rm lay}-1}+ \Phi_{\rm core}}{2} \right ] \rho_{N_{\rm lay}-1} V_{N_{\rm lay}-1} \, ,
\end{aligned}
\label{eqn:GPE_H}
\end{align}
where $\rho_i$ is the real density of layer $i$, and $V_i$ is the volume of spheroid $i$. $\Phi_i$ is the potential at the surface of spheroid $i$ and $\Phi_{\rm core}$ is the potential at the center of the body. The internal energy is given by the sum over each layer,
\begin{align}
\begin{aligned}
E_{\rm int} = & \sum_{i=0}^{N_{\rm lay}-2} \,  \rho_i \left [ V_i - V_{i+1} \right ] \epsilon \left (\rho_i, S_i \right )\\
& + \rho_{N_{\rm lay}-1} V_{N_{\rm lay}-1} \epsilon\left (\rho_{N_{\rm lay}-1}, S_{N_{\rm lay}-1} \right ) \, ,
\end{aligned}
\label{eqn:GPE_U}
\end{align}
where $\epsilon (\rho_i,S_i)$ is the specific internal energy of the material at the density and specific entropy of layer $i$, as determined by the EOS.

For this paper, we used the same HERCULES parameters as in \citet{Lock2017}. The energy components calculated using HERCULES are only weakly dependent on the number of concentric potential layers (Figure~\ref{sup:fig:HERCULES_Nlay}), the number of points used to describe potential surfaces (Figure~\ref{sup:fig:HERCULES_Nmu}), and the maximum spherical harmonic degree included in the calculation (Figure~\ref{sup:fig:HERCULES_kmax}). For the wide range of parameters we considered, each of the energy components of the body varied by less than 0.1\%. 

We considered bodies with two different thermal profiles. To calculate the properties of bodies just below the CoRoL (Stage II), we imposed thermally stratified mantles that emulate the thermal structure of bodies after an impact and during condensation \citep{Lock2017,Lock2018moon}. The core was assumed to be isentropic with a specific entropy of 1.5~kJ~K$^{-1}$~kg$^{-1}$. This core isentrope has a temperature of 3800~K at the present-day core mantle boundary (CMB) pressure, similar to the present thermal state of Earth's core \citep[e.g.,][]{Anzellini2013}. The lower mantle (75\% by mass) was assumed to be isentropic with a specific entropy of 4~kJ~K$^{-1}$~kg$^{-1}$, which is typical for the lower-mantle of calculated post-impact bodies \citep{Lock2017} (Figure~\ref{sup:fig:SpT} shows pressure-temperature profiles for forsterite isentropes). The upper 25~wt\% of the mantle was isentropic with a specific entropy of $S_{\rm outer}$ at pressures above the liquid-vapor phase boundary. At pressures below the intersection of the isentrope with the phase boundary, the body was assumed to be pure vapor on the phase boundary. This thermal profile was called a stratified structure in \citet{Lock2017}. 

For condensed bodies (in Stages III to V), we used isentropic cores and mantles with specific entropies of 1.5 and 4~kJ~K$^{-1}$~kg$^{-1}$ respectively. This mantle isentrope intersects the liquid-vapor phase boundary at low pressure (10 bar) and about 4000~K. The pressure at the surface was set at 10~bar so as not to resolve the silicate atmosphere. Our chosen thermal state approximates that of a well-mixed, liquid, magma-ocean planet with a volatile-dominated atmosphere.

The relative timings of the freezing of the mantle and tidal recession of the Moon are uncertain and so here we have used a single thermal profile for all condensed planets. However, the thermal state during tidal recession of a condensed body has little effect on the change in energy. Figure~\ref{sup:fig:mantle_entropy} shows the change in each energy term during tidal recession for bodies with mantle isentropes of 3, 3.2 and 4~kJ~K$^{-1}$~kg$^{-1}$. Mantle specific entropies of 3 and 3.2~kJ~K$^{-1}$~kg$^{-1}$ correspond to mantle potential temperatures of $\sim 1600$~K and $\sim 1900$~K, similar to the present-day and early terrestrial mantle respectively. The change in energy during tidal recession for different thermal states differs by only about a percent. 

When comparing with SPH simulations to calculate the energy change upon condensation to a magma-ocean planet (transition from stage I to stage III), we did not directly calculate the structure for all the planets. Instead, we calculated a grid of planets with a range of masses, core-mass fractions, and AM. We used total mass increments of 0.1~$M_{\rm Earth}$, core-mass fraction increments of 0.05, and a base AM increment of 0.1~$L_{\rm EM}$. When HERCULES failed to converge at the next AM step, higher AM runs were performed using a smaller step. The AM step was sequentially halved five times to provide finer AM resolution just below the CoRoL. The energy components and total energy were calculated for each body and linearly interpolated to determine the properties for a body of a given composition, mass and AM. The variation in energy for the range of parameters we considered is close to linear and this technique gives a very good approximation to the energy of directly calculated planets. 

For comparison of SPH post-impact bodies (stage I) to bodies just below the CoRoL (stage II), we interpolated a grid of hot, stratified planets with varying mass, core-mass fraction, and $S_{\rm outer}$. We ran the same mass and core-mass fraction increments as for the magma-ocean planets and $S_{\rm outer}= 4$, 4.25, 4.5, 4.75, 5.0, 5.5, 6.0, 6.5, 7.0, 7.5, and 8.0~kJ~K$^{-1}$~kg$^{-1}$, using the same AM step procedure. We found the AM of the CoRoL for each set of parameters as described in \citet{Lock2017}. The energies of a body at the CoRoL were found by linearly extrapolating from the calculated corotating planets in the same manner as in \citet{Lock2017} (see their Figure~S8). For a post-impact body with a given mass, core-mass fraction and AM, the properties of the body when it has cooled to just below the CoRoL were found by linearly interpolating the grid of CoRoL properties in mass-core fraction-AM space.

\subsection{Comparison of methods}
\label{sup:sec:methods:comparison}

\citet{Lock2017} and \citet{Lock2019pressure} have shown that, for corotating structures with a range of thermal states, the shape and pressure structure calculated by SPH and HERCULES are in good agreement. Here we examine the energy components determined using our two methods. 

Figure~\ref{sup:fig:HERCULES_comparison} shows a comparison of the energies calculated using SPH and HERCULES for Earth-mass, corotating bodies with isentropic mantles of varying specific entropy. The shape and pressure contours for these bodies are shown in Figure~4 of \citet{Lock2017}. The energies calculated using the two methods are the same to within a few percent. Given the agreement between the two different methods, we are confident in the conclusions we have made in this work.

\section{The effect of forming the Moon}
\label{sup:sec:Mooneffect}

In Figure~\ref{fig:Echange_cooling}, we neglected the effect of the formation of the Moon on the energy budget, and assumed that the magma-ocean planet had the same mass, AM and composition as the post-impact body. It is likely that little mass and AM is lost from the system during lunar accretion \citep{Ida1997,Kokubo2000,Salmon2012,Salmon2014}, but the effect of the fully-formed Moon on the energy budget must be considered. Figure~\ref{fig:Moon_effect} shows the energy difference between a system with all the mass and AM in a single magma-ocean planet and a system of a magma-ocean Earth and Moon with the Moon orbiting at a range of semi-major axes. Forming the Moon affects the energy budget in two ways. First, the Moon carries with it orbital energy. Second, the extraction of mass and AM from Earth upon forming the Moon changes its kinetic, potential and internal energy (Section~\ref{sec:results:tidal}). For bodies with initially high AM, the energy budget of Earth decreases strongly as a function of AM, the reduction in Earth's energy is dominant over the orbital energy, and the energy of the system is lower with a Moon than without. At lower AM, the orbital energy effect dominates and the energy of the total system is higher with a Moon than without. The semi-major axis of the Moon during condensation is uncertain, but, as the Moon is expected to stay close to Earth during the initial stages of cooling due to the limited tidal dissipation in a fluid body \citep[e.g.,][]{Zahnle2015}, the effect of the Moon on Earth's energy budget is likely an order of magnitude smaller than that of condensation.

\section*{References used in supplement}
\bibliographystyle{elsarticle-harv} 
\bibliography{References_manually_corrected}


\begin{table}
\caption[SPH simulation results]{Summary of SPH impact simulations used in this paper and properties of their post-impact states. For each impact, the table includes: an index number; target mass $M_{\rm t}$; number of SPH particles in target, $N_{\rm t}$; target equatorial radius, $R_{\rm t}$; target angular momentum, $L_{\rm t}$; projectile mass, $M_{\rm p}$; number of SPH particles in projectile, $N_{\rm p}$; projectile equatoiral radius, $R_{\rm p}$; projectile angular momentum, $L_{\rm p}$; impact velocity, $V_{\rm i}$; impact parameter, $b$; modified specific energy, $Q_{\rm S}$; final simulation time; bound mass of post-impact structure, $M_{\rm bnd}$; core-mass fraction of post-impact structure, $f_{\rm core}$; bound mass angular momentum, $L_{\rm bnd}$; angular velocity of the dense ($\rho$~$>$~$1000$~kg~m$^{-3}$) region of post-impact structure, $\omega_{\rho}$; spin period of dense region, $T_{\rho}$; total energy of post-impact structure, $E_{\rm tot}^{\rm bnd}$; kinetic energy of post-impact structure, $E_{\rm K}^{\rm bnd}$; potential energy of post-impact structure, $E_{\rm pot}^{\rm bnd}$; internal energy of post-impact structure, $E_{\rm int}^{\rm bnd}$; total energy of a body of the same mass, core fraction, and angular momenta as the post-impact body but with a hot partially-vaporized, thermally-stratified thermal structure with an upper mantle entropy such that the body is at the CoRoL, $E_{\rm tot}^{\rm CoRoL}$; kinetic energy at the CoRoL, $E_{\rm K}^{\rm CoRoL}$; potential energy at the CoRoL, $E_{\rm pot}^{\rm CoRoL}$; internal energy at the CoRoL, $E_{\rm int}^{\rm CoRoL}$; total energy of a magma-ocean planet with the same mass, angular momentum and core-mass fraction as the post-impact body, $E_{\rm tot}^{\rm MO}$; kinetic energy of corresponding magma-ocean planet, $E_{\rm K}^{\rm MO}$; potential energy of corresponding magma-ocean planet, $E_{\rm pot}^{\rm MO}$; internal energy of corresponding magma-ocean planet, $E_{\rm int}^{\rm MO}$; and post-impact structure dynamical class.
}
\label{sup:tab:impacts}
\end{table}

\clearpage

\begin{figure}
\centering
\includegraphics[height=0.95\textheight]{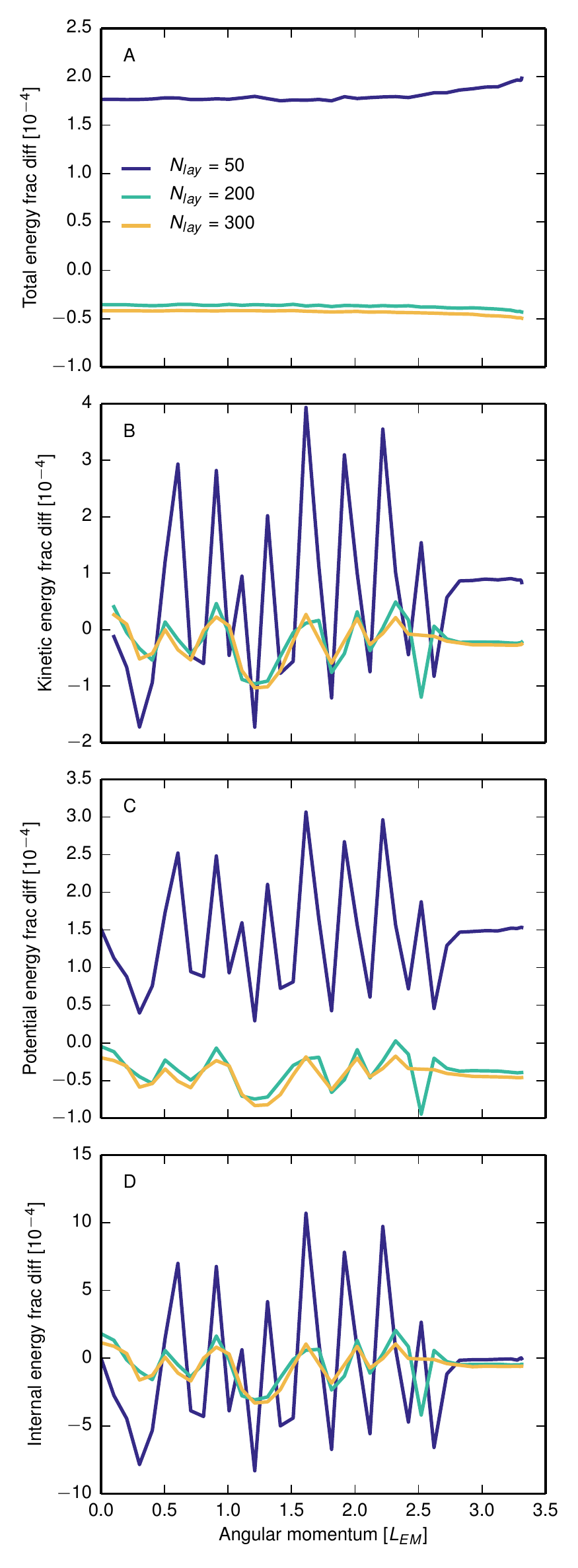}
\caption{Caption opposite.}
\end{figure}
\setcounter{figure}{0}
\begin{figure}
\caption{Energies calculated using HERCULES are only weakly dependent on the number of concentric layers used ($N_{\rm lay}$). Shown are the fractional differences in energies of Earth-mass, Earth-composition magma-ocean planets of varying angular momenta calculated using HERCULES with a given number of concentric layers (colors) and the same body calculated with $N_{\rm lay}=100$, as used elsewhere in this paper. Panels show the difference in the total (A), kinetic (B), gravitational potential (C), and internal (D) energies.
}
\label{sup:fig:HERCULES_Nlay}
\end{figure}

\begin{figure}
\centering
\includegraphics[height=0.95\textheight]{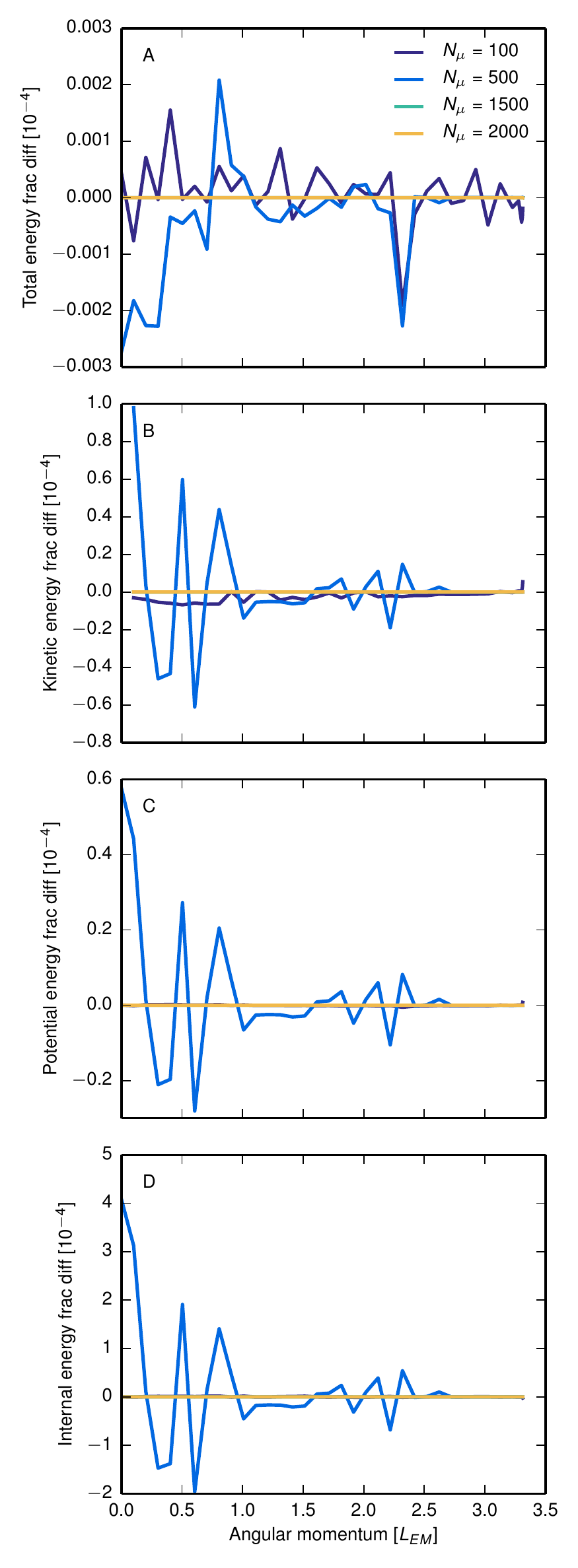}
\caption{Caption opposite.}
\end{figure}
\setcounter{figure}{1}
\begin{figure}
\caption{Energies calculated using HERCULES are only weakly dependent on the number of points used to describe each equipotential surface ($N_{\mu}$). Shown are the fractional difference in the different energy components of Earth-mass, Earth-composition magma-ocean planets calculated using HERCULES with varying $N_{\mu}$ and bodies calculated using $N_{\mu}=1000$. Panels show the difference in the total (A), kinetic (B), gravitational potential (C), and internal (D) energies.
}
\label{sup:fig:HERCULES_Nmu}
\end{figure}

\begin{figure}
\centering
\includegraphics[height=0.95\textheight]{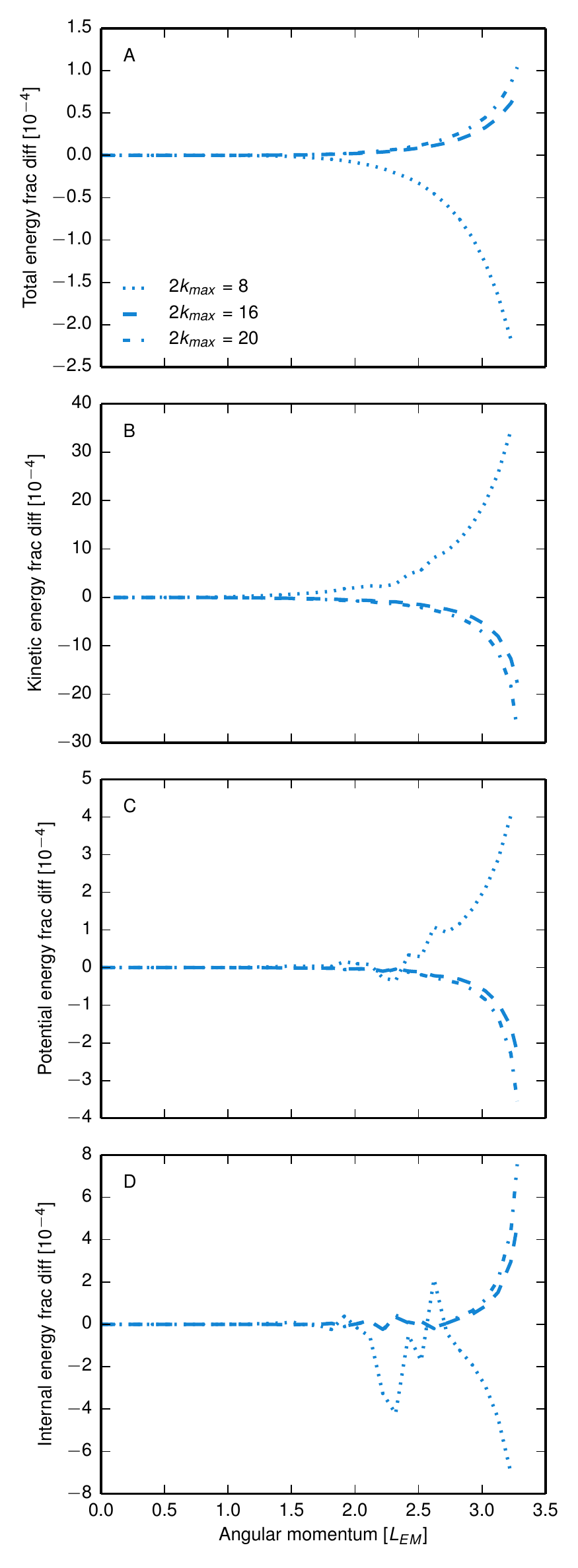}
\caption{Caption opposite.}
\end{figure}
\setcounter{figure}{2}
\begin{figure}
\caption{Energies calculated using HERCULES are only weakly dependent on the maximum spherical harmonic degree included ($2k_{\rm max}$). Shown are the fractional difference in the different energy components of Earth-mass, Earth-composition magma-ocean planets calculated using HERCULES with varying $k_{\rm max}$ and bodies calculated using $2k_{\rm max}=12$, as used elsewhere in this paper. Panels show the difference in the total (A), kinetic (B), gravitational potential (C), and internal (D) energies.
}
\label{sup:fig:HERCULES_kmax}
\end{figure}

\begin{figure}
\centering
\includegraphics{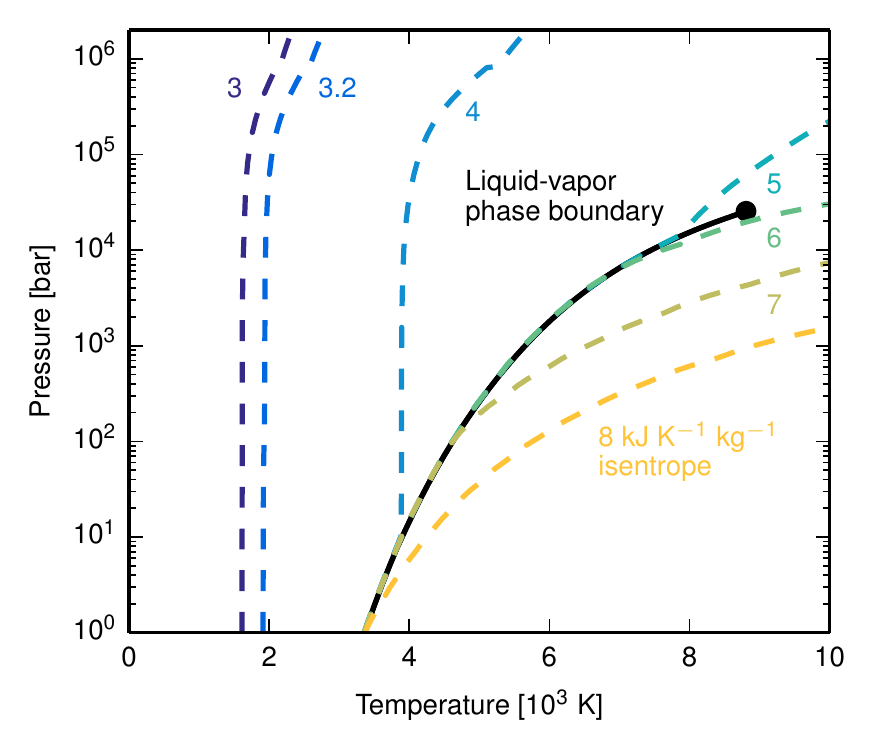}
\caption[]{Isentropes for the M-ANEOS derived forsterite EOS used in this work in pressure-temperature space. Each colored line is an isentrope for the specific entropy given by the number of the same color in kJ~K$^{-1}$~kg$^{-1}$. The black line is the liquid-vapor phase boundary. The black dot is the critical point. Adapted from \cite{Lock2017}.
}
\label{sup:fig:SpT}
\end{figure}

\begin{figure}
\centering
\includegraphics[height=0.95\textheight]{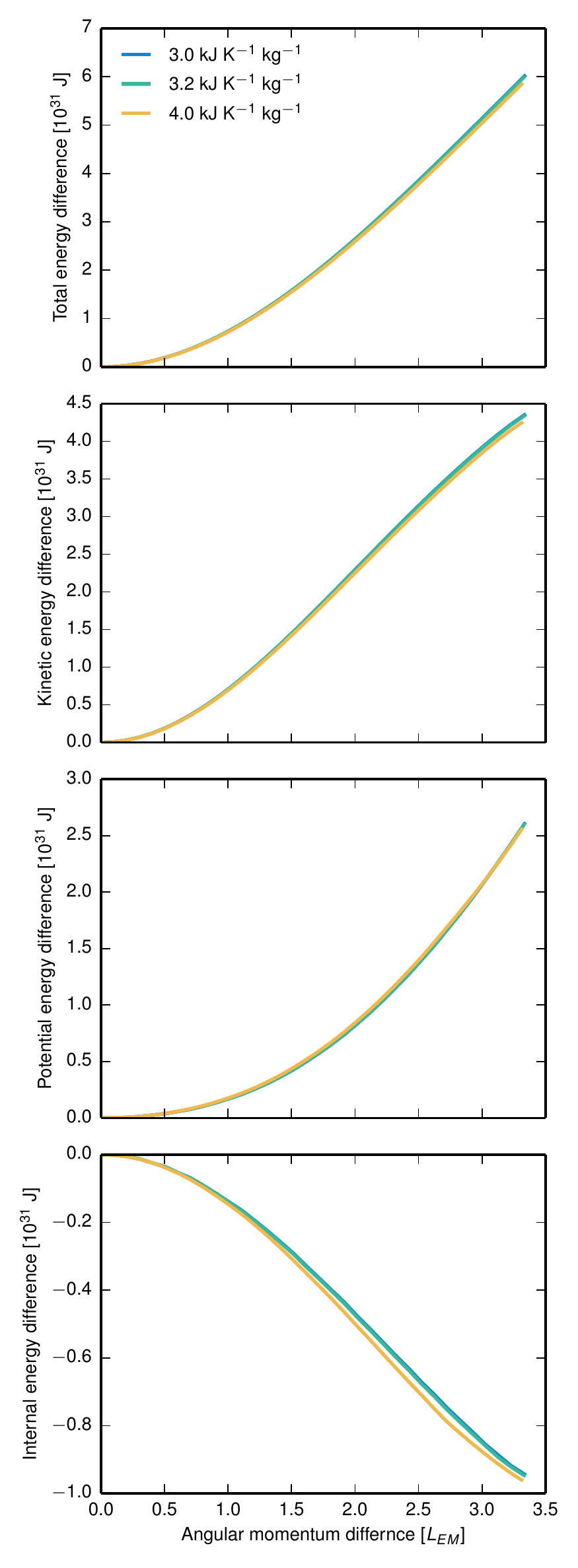}
\caption{Caption opposite.}
\end{figure}
\setcounter{figure}{4}
\begin{figure}
\caption{The change in Earth's energy budget during tidal recession is only weakly dependent on its thermal state. Shown are the differences between Earth-mass, Earth-composition bodies of given angular momenta and equivalent non-rotating bodies. Color indicate bodies with different mantle specific entropies (see Sec~\ref{sup:sec:methods:HERCULES}).
}
\label{sup:fig:mantle_entropy}
\end{figure}

\begin{sidewaysfigure*}
\centering
\includegraphics[width=\textwidth]{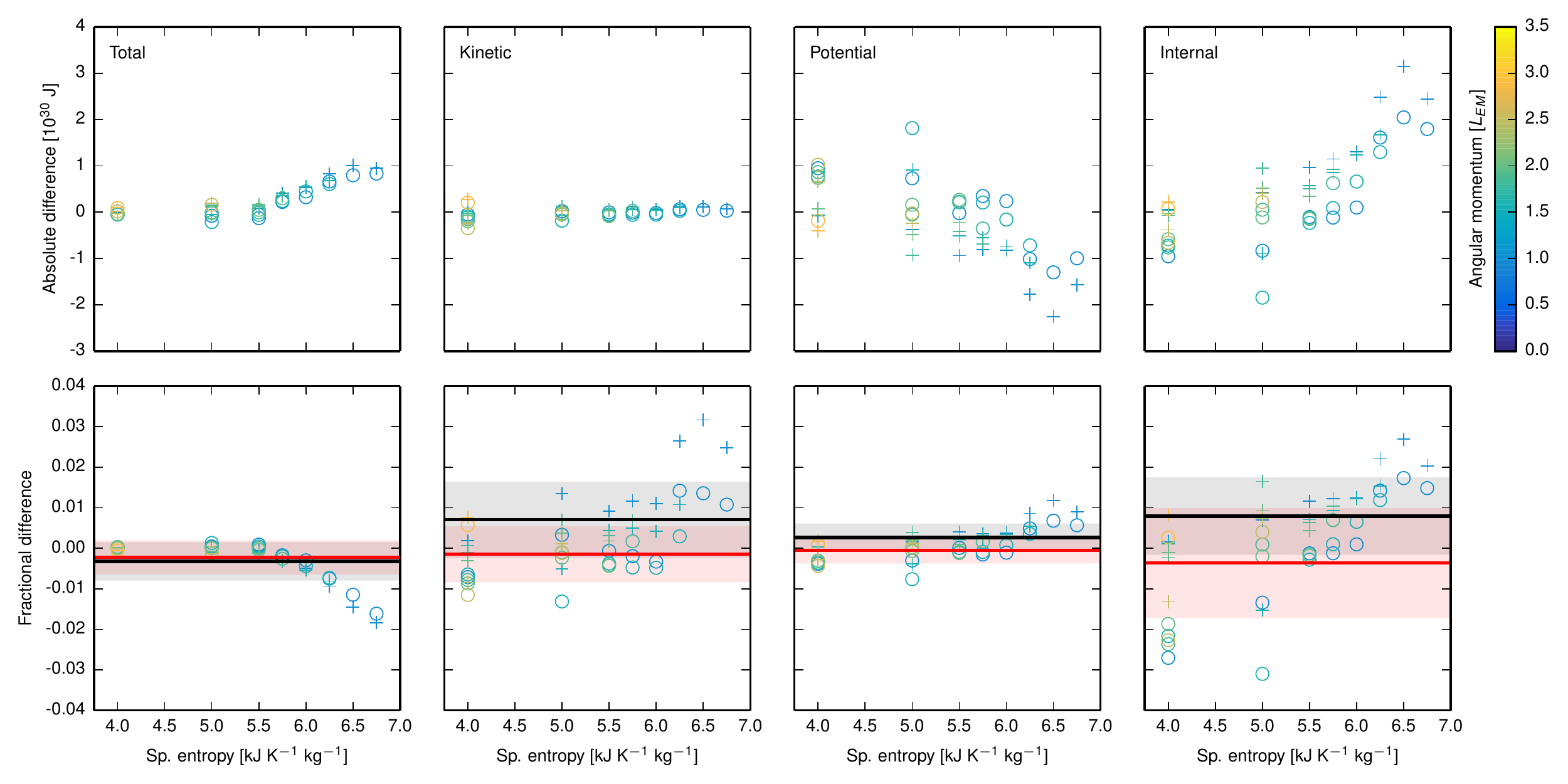}
\caption{Planetary structures calculated using SPH and HERCULES have similar energies. Rows show the absolute (top) and fractional (bottom) difference in different energy components between SPH and HERCULES calculations. Columns show the total (A), kinetic (B), gravitational potential (C), and internal (D) energies in Earth-mass bodies with different angular momenta (colors) and isentropic mantles of varying specific entropy (x-axis). Comparisons were made to HERCULES planets with a bounding pressure of both 10~bar ($+$) and a pressure equivalent to the lowest pressure in the midplane of the SPH structure ($\circ$). The solid black and red lines in the bottom row show the mean fractional difference using HERCULES planets with bounding pressures of 10~bar and the maximum SPH pressure respectively. The shaded area shows one standard deviation in the fractional errors. The shape and pressure contours for the SPH bodies plotted here are shown in Figure~4 of \citet{Lock2017}.
}
\label{sup:fig:HERCULES_comparison}
\end{sidewaysfigure*}

\begin{figure}
\centering
\includegraphics{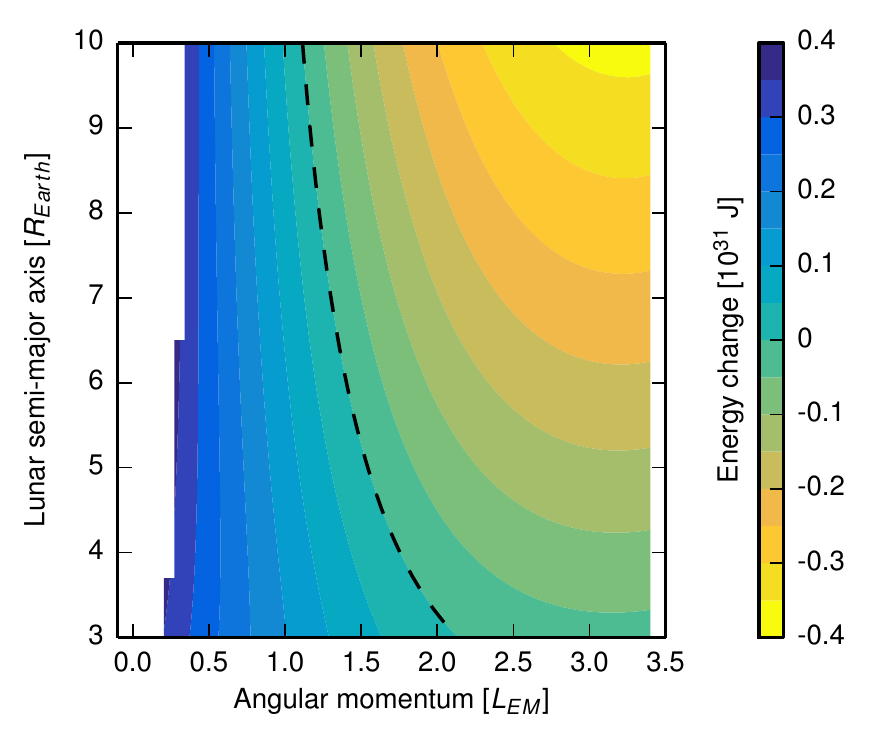}
\caption{Forming the Moon changed the energy budget but the effect was likely small compared to the effect of condensation. Shown is the difference in the total energy between an Earth-like magma-ocean planet orbited by a tidally-locked Moon at a given semi-major axis (y-axis) and a system with the same total angular momentum, mass, and composition, but with all the mass combined into a single magma-ocean planet. The orbital energy was calculated treating Earth and the Moon as point masses and the Moon was assumed to have a mantle entropy of 4~kJ~K$^{-1}$~kg$^{-1}$. The dashed line indicates the locus of points for which there is no difference in energy between the two systems.
}
\label{fig:Moon_effect}
\end{figure}


\end{document}